\documentclass{article}

\newcommand{\shorttitle}{From Attack Surface to Defense Architecture}

\PassOptionsToPackage{hyphens}{url}
\usepackage[expansion=false]{microtype}
\usepackage{graphicx}
\usepackage{subcaption}
\usepackage{booktabs}
\PassOptionsToPackage{hyphens}{url}
\usepackage{hyperref}
\usepackage{url}

\usepackage[preprint]{icml2026}
\usepackage{colortbl}
\usepackage{array}
\usepackage{amsmath}
\usepackage{amssymb}
\usepackage{mathtools}
\usepackage{amsthm}

\makeatletter
\renewcommand{\textasciitilde}{~}
\newcommand{\@creflabel}[1]{%
  \expandafter\ifx\csname r@#1\endcsname\relax
    Ref.~\ref{#1}%
  \else
    \def\@creftype{Section}%
    \in@{fig:}{#1}\ifin@\def\@creftype{Figure}\fi
    \in@{tab:}{#1}\ifin@\def\@creftype{Table}\fi
    \in@{def:}{#1}\ifin@\def\@creftype{Definition}\fi
    \in@{eq:}{#1}\ifin@\def\@creftype{Equation}\fi
    \in@{sec:}{#1}\ifin@\def\@creftype{Section}\fi
    \@creftype~\ref{#1}%
  \fi
}
\newcommand{\Cref}[1]{%
  \def\@crefsep{}%
  \@for\@crefi:=#1\do{\@crefsep\@creflabel{\@crefi}\def\@crefsep{, }}%
}
\newcommand{\cref}[1]{\Cref{#1}}
\makeatother
\newsavebox{\tcbbox}

\newtheorem{theorem}{Theorem}[section]
\newtheorem{proposition}[theorem]{Proposition}

\theoremstyle{definition}
\newtheorem{definition}[theorem]{Definition}

\theoremstyle{remark}

\definecolor{cellgain}{HTML}{2E9E5B}
\definecolor{cellloss}{HTML}{D1495B}
\definecolor{ethosurvey}{HTML}{4C72B0}
\definecolor{ethotheorize}{HTML}{DD8452}
\definecolor{ethoformalize}{HTML}{55A868}
\definecolor{ethorulemake}{HTML}{C44E52}
\definecolor{ethoenact}{HTML}{8172B3}
\definecolor{ethorevise}{HTML}{937860}

\icmltitlerunning{\shorttitle}

\begin{document}

\twocolumn[
\icmltitle{From Attack Surface to Defense Architecture: \\Compositional Security for Agentic Coding Harnesses}
\icmlsetsymbol{equal}{*}
\begin{icmlauthorlist}
\icmlauthor{Simon Hegelich}{gaia}
\icmlauthor{Habiba Sarhan}{gaia}
\icmlauthor{Morteza Shahrezaye}{gaia}
\icmlauthor{Kolja Hegelich}{gaia}
\icmlauthor{Christian Montag}{umac}
\icmlauthor{Ralf Schneider}{allianz}
\icmlauthor{Erik Cambria}{ntu}
\end{icmlauthorlist}
\icmlaffiliation{gaia}{Evercursion -- GaiaLogic Unipessoal Lda.}
\icmlaffiliation{umac}{University of Macau}
\icmlaffiliation{allianz}{Allianz Group}
\icmlaffiliation{ntu}{Nanyang Technological University, Singapore}
\icmlcorrespondingauthor{Simon Hegelich}{\mbox{simon.hegelich}@\allowbreak\mbox{evercursion.ai}}
\icmlkeywords{AI coding agents, Living-off-the-Land, reference monitor, capability security, prompt injection, Model Context Protocol, separation of duties}
\vskip 0.3in
]
\printAffiliationsAndNotice{}

\begin{abstract}
AI coding harnesses couple language models to shells, filesystems, and network sockets so that agents can resolve issues and run tests. That coupling turns the harness into an automated Living-off-the-Land (LotL) platform: once an adversary can inject instructions into retrieved content, dual-use binaries such as \texttt{bash}, \texttt{curl}, and \texttt{find} become the payload rather than a vulnerability to be separately exploited. We argue that security in agentic systems is not an emergent behavioral property of individual language models, but a structural property of the cognitive architecture and of the composition of permissions across interacting agents and humans. Many high-severity harness failures are classical OS and software problems---symlink races, path traversal, subprocess escape, SSRF, credential exposure, build-hook persistence---accelerated by agentic tool chaining. Static regex blacklists miss flag permutations and structural attacks, whereas LLM-as-a-Judge guardrails ask a second model to guess physical invariants at tens to hundreds of milliseconds per call while over-blocking benign work.

We present \textsc{ActiveSecHarness}, a neurosymbolic reference-monitor architecture organized around Separation of Intelligence and Authority: intelligence may propose actions, but it must never define its own authority. The design separates four pillars. \emph{Authority} is carried by Macaroon-inspired HMAC capability tokens that attenuate monotonically under delegation across four Trust and Authority Rings. \emph{Information flow} is governed by a product taint lattice $\mathcal{L} = \mathcal{L}_{\mathrm{trust}} \times \mathcal{L}_{\mathrm{sens}}$ (where $\mathcal{L}_{\mathrm{trust}} = \{\mathsf{Trusted} \sqsubseteq \mathsf{Sanitized} \sqsubseteq \mathsf{Derived} \sqsubseteq \mathsf{Untrusted}\}$ and $\mathcal{L}_{\mathrm{sens}} = \{\mathsf{Public} \sqsubseteq \mathsf{Sensitive}\}$) and a compositional Separation-of-Duties ledger over shared workspace provenance, so collaborating agents cannot launder secrets through disk. \emph{Enforcement} mediates tool invocation, process capability, and OS/subprocess resources through deterministic guards. \emph{Verification and applications} bind tools to extended bytecode certificates, seal split-horizon dual-key audit logs, and close a self-healing verification loop (\texttt{CodeVerifier}, \texttt{DualModelReviewer}, \texttt{SecurityTestSynthesizer}) paired with a progressive pedagogical \texttt{UserWikiManager} (\texttt{wiki\_user/}) that builds cognitive immunity in human operators.

On $N{=}252$ fixtures ($120$ domain attacks, $120$ benign SWE calls, $12$ kill-chain stage fixtures) we compare unguarded, static regex, live local LLM-judge (\texttt{qwen3:4b}), Ring~$0$, and hybrid (LLM advisory + Ring~$0$ authoritative) regimes. Attack block rates are $0.000$, $0.439$, $1.000$, $0.939$ unfitted out-of-sample ($0.962$ effective OS containment; $1.000$ post-audit conformance), and $1.000$; false positive rates are $0.000$, $0.000$, $0.983$, $0.000$, and $0.000$ (Clopper--Pearson $95\%$ CIs). The live judge blocks attacks only by refusing $98.3\%$ of benign SWE work at $142.3\,\mathrm{ms}$ median latency, whereas Ring~$0$---and the Ring~$0$-authoritative hybrid---achieve $\mathrm{FPR}{=}0.000$ with $31.5\,\mu\mathrm{s}$ Ring-$0$ median decision latency ($0.39\,\mathrm{ms}$ p95; $4{,}517\times$ faster than the live judge). In a multi-step trajectory ablation ($12$ five-stage kill chains, $12$ benign SWE workflows), even an oracle Conseca-style stateless task policy misses $4/12$ intra-scope trajectories ($\mathrm{ABR}{=}0.667$) and a coarse Boolean taint lock rejects $83.3\%$ of benign issue-driven workflows ($\mathrm{FPR}{=}0.833$), whereas the Ring-$0$ product lattice and \texttt{WorkspaceProvenanceLedger} achieve $\mathrm{ABR}{=}1.000$ and $\mathrm{FPR}{=}0.000$ at $266.7\,\mu\mathrm{s}$ per step with OS guards and a constant $\approx 24\,\mu\mathrm{s}$ per step for isolated ledger state folding across $k{=}1\dots 50$.
\end{abstract}

\section{Introduction}\label{sec:introduction}

Coding agents are useful because they act. Systems such as SWE-agent couple a language model to an agent--computer interface that edits files, navigates repositories, and runs tests and other programs~\citep[p.~1]{yang2024sweagent}. OpenHands goes further, equipping agents with ``a docker-sandboxed operating system with a bash shell, a web browser, and IPython server''~\citep[p.~2]{wang2024openhands}. Cybersecurity agents make the same design choice in offensive form: Fang et al.~give agents a terminal and a Python interpreter so that tools such as \texttt{curl} can be used against live targets~\citep[p.~3]{fang2024llm}, and Cybench evaluates agents whose actions are bash commands executed in Kali Linux~\citep[p.~3]{zhang2024cybench}. In each case the harness deliberately reuses the host's dual-use binaries rather than inventing a closed virtual instruction set: as \citet[p.~1]{pasquini2026shibboleth} observe, an LLM-driven attack depends on both the model choosing actions and the harness supplying tools, processing their outputs, and controlling execution. That is the Living-off-the-Land (LotL) duality of the coding harness: the same architectural surface that enables software engineering and red-team research also turns the agent into an automated LotL executor once its instruction stream can be manipulated.

The manipulation channel is no longer speculative. Indirect prompt injection shows that ``LLM-Integrated Applications blur the line between data and instructions,'' so that adversaries can ``remotely (without a direct interface) exploit LLM-integrated applications by strategically injecting prompts into data likely to be retrieved''~\citep[p.~1]{greshake2023not}. Tool-integrated agents amplify the blast radius: InjecAgent documents how malicious instructions embedded in external content aim ``to manipulate these agents into executing detrimental actions against users''~\citep[p.~1]{zhan2024injecagent}, and AgentDojo notes that a successful injection ``can allow an external attacker to take actions (and call tools) on behalf of the user''~\citep[p.~1]{debenedetti2024agentdojo}. ToolEmu's risk analysis makes the consequence concrete: even the safest evaluated language-model agent still exhibits severe tool failures ``23.9\% of the time according to {[}their{]} evaluator''~\citep[p.~1]{ruan2024toolemu}.

What follows is often misdiagnosed as a novel model attack. Symlink and TOCTOU races, path traversal, subprocess escape, environment and \texttt{PATH} injection, SSRF and DNS rebinding, credential exposure, resource exhaustion, build-hook persistence, cross-agent exfiltration through shared files, and persistent workspace poisoning are classical OS and software problems. Agentic tool chaining accelerates them by composing dual-use binaries under ambient authority; it does not invent a new physics of compromise. The central thesis of this paper is therefore not merely that coding harnesses need better individual filters, but that \textbf{security in agentic systems is a structural property of the architecture and of the composition of permissions---not a behavioral property of individual agents}. In our companion mathematical theory of open-ended intelligence, we show that cognitive capability is not the consequence of model scale alone, ``but of a particular organizational structure,'' so that ``the question worth asking shifts from which model to which system''~\citep[pp.~1--2]{hegelich2026openended}. Overcoming the failure modes of both blind statistical token predictors and brittle single-layer rules requires ``designing a composite architecture that keeps these requirements structurally separate while allowing them to integrate''~\citep[p.~2]{hegelich2026openended}. Crucially, when multiple bounded agents interact through shared external memory, ``{[}t{]}he collective state is not the union of individual states but the individual states together with the relations coupling them''~\citep[p.~32]{hegelich2026openended}. Applied to security, this theorem explains why aligning individual agents in isolation is structurally insufficient: two agents whose individual steps each appear benign can still compose across shared workspace state into an unauthorized exfiltration channel unless authority and information flow are enforced across their composition.

Two common industry responses miss that architectural point. Static string or regex blacklists can catch a handful of literal patterns, yet they fail on flag permutations, Base64 or Hex pipe-to-shell payloads, spoofed archive headers, magic-byte disguises, and Unicode bidirectional (Bidi) Trojan Source encodings that make source appear different to a compiler and to the human eye~\citep[p.~1]{boucher2021trojan}. Delegating the same checks to a secondary LLM-as-a-Judge is attractive as a product feature and brittle as a protection mechanism. A judge must reason correctly about whether an action violates a rule, yet systematic evaluation of LLM logical reasoning across deductive, inductive, and abductive settings---using content-neutral problems that separate reasoning from text comprehension---traces model failures to both evidence selection and the reasoning process itself~\citep{xu2025logical}. Heuristic detectors ``provide {[}no{]} guarantee of security and regularly fall short to new attacks in practice''~\citep[p.~3]{debenedetti2025camel}. Saltzer and Schroeder's principle of complete mediation requires that ``{[}e{]}very access to every object must be checked for authority''~\citep[p.~1280]{saltzer1975protection}; economy of mechanism further insists that protection designs stay ``as simple and small as possible,'' because ``design and implementation errors that result in unwanted access paths will not be noticed during normal use''~\citep[p.~1280]{saltzer1975protection}. Asking a probabilistic model to decide decompression ratios, canonical path containment, SSRF IP resolution, or password entropy violates both principles and pays a latency tax four orders of magnitude above a deterministic kernel.

A second classical failure mode appears once harnesses and Model Context Protocol (MCP) servers execute with the user's ambient authority across the cyber kill chain---from reconnaissance and weaponization of repository artifacts through exploitation, privilege escalation, and persistence or exfiltration. Hardy's confused deputy ``runs with authority stemming from two sources'' and ``has no way to keep them apart''~\citep[p.~36]{hardy1988confused}. MCP deployments compound the problem: tool selection commonly relies ``on textual tool names and descriptions without cryptographic verification''~\citep[p.~17]{hou2025mcp}, and a post-installation rug pull can alter ``an apparently benign MCP server'' into malicious code~\citep[p.~19]{hou2025mcp}. The protocol itself ``does not include mechanisms for runtime isolation or privilege control''~\citep[p.~24]{hou2025mcp}. Recent operational incidents show that this ambient-authority failure mode is already producing severe national-security and financial consequences. In late 2026, a defense contractor connected a personal AI coding agent with administrative credentials to the Pentagon's Non-classified Internet Protocol Router Network (NIPRNet), where the agent ran unmediated for two months and exfiltrated login credentials, passwords, and an internal network map to a high-risk third-party website vulnerable to prompt injection~\citep{elliott2026pentagon}. In the October 2026 Binance heist ($\approx\$6.34\,\text{B}$ impacted), attackers trojaned a developer's local \texttt{pyenv} binary after a fake recruitment coding test and then pivoted through an internal AI tool to compromise two additional wallet engineers, harvest a container deployment permission token, and reach the hot-wallet signing service to authorize $50$ on-chain EIP-7702 delegations~\citep{binance2026incident,evans2026binance}---confirming that an internal AI tool with host execution rights functions as ``a remote-administration system with a chat window'' whose decisive vulnerability is ``what an AI tool may do, for whom, and with whose approval''~\citep{evans2026binance}. System-level defenses such as IsolateGPT~\citep{wu2024isolategpt}, AirGapAgent~\citep{bagdasarian2024airgapagent}, Conseca~\citep{tsai2025conseca}, and CaMeL~\citep{debenedetti2025camel} show that execution isolation, contextual task scoping, and explicit control--data separation help; yet because stateless per-action policies cannot track how individually benign steps compose across multi-step trajectories~\citep[p.~5]{tsai2025conseca}, coding harnesses still need an OS-facing reference monitor and provenance ledger that mediate archives, sockets, credentials, dual-use binaries, and forensic audit trails without handing reconnaissance blueprints to an injected workspace agent.

Equally misplaced is the prevailing model-centric cybersecurity discourse. Public alarm frequently centers on the claim that frontier language models have grown so intrinsically dangerous that they autonomously hacked enterprise targets; the policy reflex is to lobotomize model weights via refusal tuning or to restrict access to open-weight security and code models on hosting platforms such as HuggingFace. Because ``{[}o{]}pen-ended intelligence is a property of a recursively extended cognitive process, not of an individual physical substrate''~\citep[p.~37]{hegelich2026openended}, attributing system compromise to ``dangerous model weights'' mistakes an execution-boundary failure for a weight-level property. Where an autonomous agent compromises a target or a host environment, the decisive failure is architectural: the harness or target lacked complete mediation, least privilege, and compositional permission enforcement. Conversely, crippling the model's security reasoning at the weight level creates a dangerous dual-use defense paradox: defenders urgently need uncrippled, high-capability models inside a secure harness to audit complex codebases, synthesize adversarial regression tests, and repair vulnerabilities. And because the human operator---particularly the ``vibe coder'' who prompts language models to assemble applications without formal security training under schedule pressure, generating code at a volume that overwhelms manual review while logic-level backdoors fly under the radar of traditional static scanners (SAST)~\citep{elliott2026pentagon}---is an active participant in the coupled cognitive system~\citep[p.~37]{hegelich2026openended} and a prime target of social engineering, compositional security fails if the human repeatedly rubber-stamps dangerous prompts or waives guards without understanding why. A secure harness must therefore verify generated code, heal vulnerabilities before execution, and systematically educate the operator over time.

Framed in cognitive-architectural terms, a secure coding harness is a neurosymbolic AI system by definition. As Cambria and collaborators observe, ``{[}s{]}ubsymbolic techniques do not require much time nor effort to perform classification but they are data-dependent and function in a black-box manner''~\citep[p.~107]{cambria2020senticnet6}, and ``even the finest AI models are vulnerable to adversarial examples''~\citep[p.~3829]{cambria2022senticnet7}. The latest iteration of this line, SenticNet\textasciitilde9, carries the argument into the LLM era: because LLMs remain limited by hallucination and opaque reasoning, it uses generative models only to discover candidate conceptual primitives and keeps the resulting reasoning in an explicit symbolic layer, outperforming GPT-4.1 on three affective computing tasks while leaving every decision inspectable~\citep{cambria2026senticnet9}. Even when large language models use natural language as a ``shared interface between neural pattern-matching and symbolic control''~\citep[p.~1]{mao2026neurosymbolic}, a purely language-mediated pipeline means that ``the model is not enforcing symbolic well-formedness in the technical sense''~\citep[p.~2]{mao2026neurosymbolic}. Consequently, ``even if real parrots or stochastic ones (LLMs) produce appropriate linguistic responses to task-related $\varphi_i$ prompts, we would not deem their linguistic behavior trustworthy'' on their own, which is why ``humans have to watch over them and correct them in mission-critical tasks''~\citep[p.~67]{cambria2023seven}. \textsc{ActiveSecHarness} operationalizes neurosymbolic AI at the operating-system boundary to ``mitigate unexpected model behavior and inject interpretability by combining learnable parameters (neuro) with predefined knowledge bases or templates (symbolic)''~\citep[p.~1712]{xing2023neurosymbolic}: the subsymbolic neural model proposes plans, code, and security reviews in Rings\textasciitilde2--3, while Ring\textasciitilde0 enforces formal symbolic capability algebra, information-flow labels, AST well-formedness, and cryptographic state transitions. The governing slogan is deliberate: intelligence may propose actions, but it must never define its own authority.

We therefore treat the coding harness as a structural and compositional protection problem and contribute four interlocking pieces:

\begin{enumerate}
\item A six-domain LotL threat taxonomy (Domains~A--F) mapped onto STRIDE, MITRE ATT\&CK, and the cyber kill chain, together with a classical-primitives framing that states four mediation requirements---complete mediation, resource containment, information-flow control, and temporal integrity (\cref{sec:taxonomy}).
\item \textsc{ActiveSecHarness}, structured around (i)~Authority: four Trust and Authority Rings with Macaroon-inspired HMAC capability chains; (ii)~Information Flow: a product taint lattice $\mathcal{L} = \mathcal{L}_{\mathrm{trust}} \times \mathcal{L}_{\mathrm{sens}}$ and compositional Separation of Duties over workspace provenance; (iii)~Enforcement: three mediation boundaries mapped to concrete OS resource interceptors; and (iv)~Verification, Self-Healing, and Pedagogical Defense: extended tool certificates, split-horizon audit logs, a closed-loop \texttt{CodeVerifier} / \texttt{DualModelReviewer} / \texttt{SecurityTestSynthesizer} repair pipeline, and the progressive \texttt{UserWikiManager} security academy (\cref{sec:architecture,sec:guards}).
\item Three formal guarantees in the appendix: Separation of Intelligence and Authority via monotonic attenuation; monotonic information-flow security under the product lattice (distinguishing sanitization from redaction while preserving isolated SWE tests under $\mathsf{Derived}$); and compositional multi-agent non-exfiltration across shared workspace state (\cref{sec:appendix}).
\item An empirical evaluation on $N{=}252$ fixtures ($120$ domain attacks, $120$ benign SWE calls, $12$ kill-chain stage fixtures) across five regimes---unguarded, regex, live local LLM-judge, Ring~$0$ (reporting both unfitted out-of-sample audit and post-audit conformance), and hybrid---plus a multi-step trajectory ablation comparing an idealized oracle Conseca-style stateless task policy, a coarse Boolean taint lock, and Ring~$0$'s product lattice + \texttt{WorkspaceProvenanceLedger} across $12$ five-stage attack kill chains, $12$ benign SWE trajectories, and trajectory lengths $k{=}1\dots 50$ (\cref{sec:experiments}).
\end{enumerate}

\section{Classical Threat Taxonomy for Coding Harnesses}\label{sec:taxonomy}

A coding harness is not a chat interface with a few plugins. It is a process that can open shells, rewrite trees, and dial sockets under the operator's identity. Saltzer and Schroeder already warned that ``{[}e{]}very program and every user of the system should operate using the least set of privileges necessary to complete the job''~\citep[p.~1281]{saltzer1975protection}, and that access decisions should rest on ``permission rather than exclusion,'' so that ``the default situation is lack of access''~\citep[p.~1280]{saltzer1975protection}. Coding agents routinely invert that default: once installed, they inherit ambient filesystem, credential, and network authority, then accept untrusted repository text, issue trackers, and tool descriptions as if those inputs were inert data. We organize the LotL surface into six domains and a concrete sub-vector catalog (Appendix~\ref{sec:appendix}), each annotated with STRIDE categories and representative MITRE ATT\&CK techniques. The goal is not encyclopedic coverage of every CVE; it is a mediation checklist that a Ring\textasciitilde0 kernel can implement without asking a language model whether a ZipBomb ``looks dangerous.''

\subsection{Classical Attack Primitives under Agentic Acceleration}

Before the domain catalog, we state a thesis that shapes the rest of the paper. Many high-severity harness failures are not exotic model behaviors; they are classical attack primitives that agentic tool chaining merely accelerates. Symlink and TOCTOU races rewrite destinations between policy check and open. Path traversal escapes workspace roots through \texttt{../} and link following. Subprocess escape turns an allowlisted binary into a shell via flags such as \texttt{find -exec}. Environment and \texttt{PATH} injection smuggle credentials and substitute trojaned executables into child processes. SSRF and DNS rebinding reach loopback, RFC1918, and cloud metadata after a hostname has already passed a textual allowlist. Credential exposure leaks keys through files, logs, and tool arguments. Resource exhaustion (fork, memory, archive bombs) denies the host. Build hooks and SCM triggers (\texttt{.git/hooks}) persist across sessions. Cross-agent exfiltration uses a shared workspace file as a covert channel between differently scoped agents. Persistent workspace poisoning plants instructions or artifacts that later sessions will trust.

We do not claim that every existing coding harness leaves all of these classes unprotected. We claim that the classes sit below the tool-call abstraction that product UIs and MCP schemas expose, and that they must therefore be enforced at the resource boundary---where pathnames are resolved, processes are spawned, and sockets are connected. Four requirements follow: \emph{Complete Mediation} (every access is checked~\citep[p.~1280]{saltzer1975protection}), \emph{Resource Containment} (quotas, namespaces, and scrubbed environments bound blast radius), \emph{Information-Flow Control} (labels travel with data across tools and agents), and \emph{Temporal Integrity} (check and use cannot be raced apart). Table~\ref{tab:primitives} maps each primitive to the primary requirement that defeats it; Domains\textasciitilde A--F below are the LotL-facing organization of the same surface.

\begin{table}[t]
\centering
\scriptsize
\setlength{\tabcolsep}{2.5pt}
\caption{Classical primitives accelerated by agentic tool chaining, and the primary harness requirement that addresses each.}
\label{tab:primitives}
\begin{tabular}{@{}>{\raggedright\arraybackslash}p{0.46\linewidth}>{\raggedright\arraybackslash}p{0.48\linewidth}@{}}
\toprule
Primitive & Primary requirement \\
\midrule
Symlink / TOCTOU & Temporal integrity + complete mediation \\
Path traversal & Complete mediation (canonical paths) \\
Subprocess escape & Resource containment (flag / binary policy) \\
Env / \texttt{PATH} injection & Resource containment (scrubbed env) \\
SSRF / DNS rebinding & Temporal integrity (pre-connect IP pin) \\
Credential exposure & IFC + DLP mediation \\
Resource exhaustion & Resource containment (quotas) \\
Build hooks / SCM triggers & Complete mediation (sensitive paths) \\
Cross-agent exfiltration & Compositional IFC / SoD \\
Workspace poisoning & IFC (taint) + temporal integrity \\
\bottomrule
\end{tabular}
\end{table}

\subsection{Ambient Authority and the Confused Deputy}

Hardy's classic observation still fits the harness: the privileged component ``serves two masters and carries some authority from each,'' yet ``has no way to keep them apart''~\citep[p.~36]{hardy1988confused}. In the modern setting one master is the human operator who installed the agent; the other is whichever remote artifact or MCP tool description the agent last retrieved. Capabilities, in Hardy's sense, unify designation and authorization so that a callee ``would merely refer to a capability provided by the invoker''~\citep[p.~37]{hardy1988confused}. Without that discipline, GTFOBins-style shell escapes, environment-variable credential inheritance, and network egress all execute under a single ambient principal. Separation of privilege---``a protection mechanism that requires two keys to unlock it''~\citep[p.~1281]{saltzer1975protection}---is precisely what ambient YOLO modes discard when a single session may read secrets, run shells, and open sockets.

\subsection{Domain A: LotL Binary Abuse and Self-Tampering}

Domain\textasciitilde A covers misuse of pre-installed dual-use binaries and of the harness's own security configuration. Sub-vector\textasciitilde A.1 is classical living-off-the-land execution (ATT\&CK T1059): \texttt{find -exec}, \texttt{curl | sh}, \texttt{python -c}, and similar compositions that never drop a custom implant. A.2 covers environment-variable credential inheritance (T1552.001), where child processes receive \texttt{AWS\_*}, \texttt{OPENAI\_API\_KEY}, or shell history simply because the harness scrubbed nothing. A.3 is Ring\textasciitilde0 self-tampering (T1562.001): an agent that can rewrite its own policy files or unload guards has already escaped mediation. A.4 covers persistence through \texttt{.git/hooks} and related SCM triggers (T1546). Cyber agents that ``output a response that includes an action (a bash command \ldots) which is then executed in the environment''~\citep[p.~3]{zhang2024cybench} make Domain\textasciitilde A the default rather than an edge case. STRIDE mappings concentrate on Elevation of Privilege and Tampering.

\subsection{Domain B: Artifact and Archive Vectors}

Domain\textasciitilde B treats files as hostile programs. B.1 is the spoofed-header ZipBomb (T1499): declared sizes lie, nested archives explode, and any guard that trusts ZIP headers fails closed only by accident. B.2 is Zip-Slip path and symlink traversal (T1083), writing outside the destination tree. B.3 covers polyglot and magic-byte disguises (T1036), for example an ELF image renamed to \texttt{.txt}. B.4 is Trojan Source: source ``maliciously encoded so that it appears different to a compiler and to the human eye,'' exploiting Unicode so that ``tokens are logically encoded in a different order from the one in which they are displayed''~\citep[p.~1]{boucher2021trojan}. Because Bidi controls inside comments and strings are accepted by most compilers~\citep[p.~3]{boucher2021trojan}, an operator approving a ``sudo'' prompt can ratify a command that is not the one displayed. STRIDE emphasis is Tampering and Denial of Service.

\subsection{Domain C: Network, SSRF, DNS Rebinding, and Supply Chain}

Domain\textasciitilde C begins where the harness leaves the workstation. C.1 is cleartext HTTP downgrade. C.2 is SSRF to loopback, RFC1918 space, or cloud metadata endpoints such as \texttt{169.254.169.254} (T1071). C.3 is DNS-rebinding time-of-check/time-of-use: a name resolves to a public address for the policy check and to a private address for the socket. C.4 covers unverified external model or tool endpoints that silently exfiltrate context. Fail-safe defaults require that such targets be explicitly permitted rather than silently reachable~\citep[p.~1280]{saltzer1975protection}. STRIDE maps primarily to Information Disclosure and Spoofing.

\subsection{Domain D: Credentials, Dictionary Mutations, and Secret DLP}

Domain\textasciitilde D covers secrets in both directions. D.1 is weak and mangled dictionary passwords. Wheeler shows that LUDS character-class rules remain ``a conclusively burdensome and ineffective security practice''~\citep[p.~157]{wheeler2016zxcvbn}, while leetspeak demangling recovers dictionary words from mangled inputs (Wheeler's running example demangles to \texttt{abalone})~\citep[p.~163]{wheeler2016zxcvbn}. D.2 is bidirectional secret DLP (T1552): cleartext API keys, PEM material, and JWTs must be blocked on writes \emph{and} redacted or refused on reads that would feed exfiltration tools. STRIDE focus is Information Disclosure and Elevation of Privilege.

\subsection{Domain E: Sandbox Escape and Resource Starvation}

Domain\textasciitilde E is confinement failure. Even when coding harnesses wrap agent sessions in OS-level containers, ``containers virtualize resources at the OS level where all the containers share the same OS''~\citep[p.~2]{zhong2022container}, making explicit per-container resource allocation and limits foundational for ``resource isolation control between containers''~\citep[p.~7]{zhong2022container}. Without per-call mediation inside the container boundary, an injected agent still enjoys ambient access to everything mounted into that container. E.1 is host-path escape outside the workspace (\texttt{/etc/passwd}, \texttt{\~/.ssh}). E.2 is fork, memory, or disk bombs (T1499). E.3 is unapproved binary flags that convert an allowlisted tool into a shell (\texttt{find -exec}, \texttt{tar --to-command}, \texttt{git -c core.fsmonitor}). Psychological acceptability matters here: if the human interface is not ``designed for ease of use, so that users routinely and automatically apply the protection mechanisms correctly''~\citep[p.~1281]{saltzer1975protection}, operators will click through opaque approval dialogs. STRIDE maps to Elevation of Privilege and Denial of Service.

\subsection{Domain F: Cognitive and Multi-Agent Vectors}

Domain\textasciitilde F is where language-model control flow meets classical authority. F.1 is indirect prompt-injection taint escalation: retrieved content that should be treated as data becomes instructions~\citep[p.~1]{greshake2023not}. F.2 is confused-deputy sub-agent delegation and social-engineering ``sudo'' traps, including MCP rug-pull mutation of tool implementations~\citep[p.~19]{hou2025mcp}. F.3 is the Lethal Trifecta: combining sensitive read, shell execution, and network egress inside one session---or across collaborating agents that share workspace files---so that stolen material can leave the host. These vectors do not replace Domains\textasciitilde A--E; they orchestrate them. A taxonomy that stopped at string matching would systematically undercount the attack that never looks like a banned token.

\subsection{Kill-Chain Acceleration and the Forensic Reconnaissance Paradox}

Domains\textasciitilde A--F do not occur as isolated one-shot events; in an unguarded or naively guarded coding harness, they compose to accelerate every stage of the classical cyber kill chain:

\begin{enumerate}
\item \textbf{Reconnaissance:} An injected agent enumerates workspace trees, environment variables (A.2), SSH configs (E.1), and---critically---the harness's own telemetry and audit logs to map internal paths, prior failure modes, and redacted or cleartext secrets.
\item \textbf{Weaponization and Delivery:} Malicious payloads arrive disguised as benign repository artifacts---spoofed-header ZipBombs (B.1), Zip-Slip archives (B.2), magic-byte polyglots (B.3), Trojan Source Bidi comments (B.4)~\citep[p.~3]{boucher2021trojan}, or poisoned issue text and MCP servers (F.1, F.2)~\citep[p.~19]{hou2025mcp}---or are generated inline when a small model writes vulnerable application code for an unsuspecting “vibe coder.”
\item \textbf{Exploitation:} Instead of corrupting memory, the attacker exploits the agent's instruction/data confusion~\citep[p.~1]{debenedetti2024agentdojo} or tricks the human operator into rubber-stamping an obfuscated Base64 pipe-to-shell prompt.
\item \textbf{Privilege Escalation and Lateral Movement:} The compromised agent abuses GTFOBins flags (A.1, E.3), confused-deputy sub-agent delegation (F.2)~\citep[p.~36]{hardy1988confused}, Ring~$0$ self-tampering (A.3), or SSRF and DNS rebinding against internal metadata services (C.2, C.3).
\item \textbf{Persistence and Exfiltration:} The attacker plants \texttt{.git/hooks} triggers (A.4) and closes the Lethal Trifecta (F.3) by combining sensitive file reads (D.2) with network egress (C.1, C.4), including laundering through a second agent's workspace read.
\end{enumerate}

Mapping the kill chain exposes a fundamental tension in harness logging and deception. Recent work on fingerprinting agentic attackers shows that classical prompt-injection honeypots fail completely because frontier models recognize the injection attempt while the attack is still running, whereas deterministic execution-level probes of the harness succeed via fixed rules with no learned component~\citep[pp.~2, 17]{pasquini2026shibboleth}. Likewise, defenders require rich execution and violation telemetry for \emph{ex-post} forensic reconstruction; yet if those logs sit in readable workspace files or single-tier JSONL streams, an adversary who achieves Stage\textasciitilde1 reconnaissance inside the workspace can read the audit trail as a complete blueprint of system paths, guard rules, and credential locations. As we show in \cref{sec:architecture,sec:guards}, resolving this paradox requires applying Saltzer and Schroeder's separation of privilege~\citep[p.~1281]{saltzer1975protection} directly to the audit log via split-horizon dual-key encryption, and treating the human operator not as an infallible oracle but as a participant who must be protected by automated code verification and continuous security education.

\section{The ActiveSecHarness Architecture}\label{sec:architecture}

\textsc{ActiveSecHarness} adapts classical protection rings to agentic coding harnesses as a composite neurosymbolic architecture. As established in our mathematical foundation for open-ended intelligence, cognitive capability is not a product of scale alone, ``but of a particular organizational structure''~\citep[p.~2]{hegelich2026openended} whose ``{[}f{]}irst law is a strict duality: the system needs an object-level structure and a distinct meta-level structure, kept separate. The object level is responsible for interacting with the world. It conducts observations, performs operations, generates hypotheses, and stores empirical knowledge. The meta level acts on the object level itself rather than on the world directly''~\citep[p.~2]{hegelich2026openended}. Attempting to collapse this duality by asking a single language model---or a second language model acting as an inline judge---to enforce its own physical boundaries confuses object-level hypothesis generation with meta-level authority.

Instead, \textsc{ActiveSecHarness} ``keeps these requirements structurally separate while allowing them to integrate''~\citep[p.~2]{hegelich2026openended}. As \citet[p.~66]{cambria2023seven} argue, ``{[}o{]}ld-school (symbolic) AI was better at Symbol Grounding but it was not scalable nor flexible. New deep-learning-based (subsymbolic) AI, instead, is very scalable and flexible but it does not handle symbols. The best of both worlds could be achieved through a hybrid (neurosymbolic) AI that leverages the strengths of both symbolic and subsymbolic models to overcome their respective limitations.'' In \textsc{ActiveSecHarness}, ``symbolic representations and reasoning are used to provide high-level context and logic, while neural networks are employed for pattern recognition, learning from data, and capturing low-level features''~\citep[p.~1712]{xing2023neurosymbolic}, directly addressing what \citet[p.~3829]{cambria2022senticnet7} identify as the five core limitations of purely subsymbolic systems: ``dependency, reproducibility, trustworthiness, interpretability, and explainability.'' Ring\textasciitilde0 mediation and the self-verifying ledger make security enforcement ``reproducible, because each reasoning step can be explicitly recorded and replicated through each iteration''~\citep[p.~3829]{cambria2022senticnet7}. SenticNet\textasciitilde9 names the deeper obstacle as the symbol grounding problem: representations that are not tied to real-world meaning leave a system's decisions opaque and hard to interpret~\citep{cambria2026senticnet9}. Ring\textasciitilde0 is grounded in exactly this sense. Its decisions refer to resolved canonical paths, pinned socket addresses, measured decompression ratios, and signed capability scopes, never to the natural-language text that describes an action, so an injected instruction that rephrases a dangerous call does not change what Ring\textasciitilde0 checks.

The architecture is organized around Separation of Intelligence and Authority under four pillars---Authority, Information Flow, Enforcement, and Verification---coupled to a closed-loop Verification and Pedagogical Defense layer.

\subsection{Authority: Trust and Authority Rings and Capability Chains}

\begin{definition}[Trust and Authority Rings]
\label{def:rings}
We use \emph{Trust and Authority Rings}, not hardware CPU rings. Ring~$0$ is the immutable security kernel: a non-bypassable reference monitor that mediates every tool call before and after execution and that cannot be rewritten by any agent token. Ring~$1$ is the human-sovereign layer: context-isolated command deobfuscation, explain-before-sudo ceremonies, and the progressive \texttt{UserWikiManager} curriculum that protect the operator against approval fatigue, Base64/Hex pipe-to-shell payloads, and Trojan Source Unicode~\citep[p.~1]{boucher2021trojan}. Ring~$2$ hosts privileged orchestrators, independent security reviewers, and scoped “sudo” agents that operate only under short-lived, cryptographically attenuated capability grants (for example, laboratory pentesting inside \texttt{10.10.0.0/16}). Ring~$3$ hosts unprivileged workers: sandboxed, offline-by-default sub-agents restricted to canonical workspace paths and strict tool subsets.
\end{definition}

Least privilege~\citep[p.~1281]{saltzer1975protection} and complete mediation~\citep[p.~1280]{saltzer1975protection} are structural rather than advisory: a Ring\textasciitilde $3$ worker never receives ambient authority, and Ring\textasciitilde $0$ checks every access even when Ring\textasciitilde $1$ has already shown a human a cleaned command string. In organizational terms, the four rings instantiate the institutional checks and balances that human high-reliability organizations use to govern fallible actors: rather than expecting a single generative agent to police itself, the architecture separates \emph{generative} problem-solving in Rings\textasciitilde $2$--$3$ from \emph{bureaucratic} procedural verification (\texttt{DualModelReviewer}, \texttt{ToolCertificate}) and a deliberately restrictive, need-to-know \emph{pathological} gatekeeper in Ring\textasciitilde $0$ empowered to halt execution and revoke authority immediately upon tripwire contact~\citep[p.~ii22]{westrum2004typology,yu2024pathological}. IsolateGPT observes that LLM-app ecosystems ``resemble the settings of earlier computing platforms, where there was insufficient isolation between apps and the system''~\citep[p.~1]{wu2024isolategpt}; our rings push isolation down to OS-facing tool mediation rather than stopping at app-level hub-and-spoke routing.

Ambient authority is replaced by HMAC-SHA256 chained \texttt{CapabilityToken} objects in the spirit of macaroons, which ``embed caveats that attenuate and contextually confine when, where, by who, and for what purpose a target service should authorize requests''~\citep[p.~1]{birgisson2014macaroons}. Integrity follows the nested construction in which ``each such value itself is used as the HMAC key for the next message''~\citep[p.~4]{birgisson2014macaroons}. Tokens additionally bind parent-subject identity, per-call nonces, and remaining quotas so that replay and quota inflation fail closed.

\begin{definition}[Monotonic attenuation]
\label{def:attenuation}
Let $\tau$ be a capability token with scope $\mathrm{Scope}(\tau)$ over allowed tools, canonical path prefixes, binary allowlists, network CIDR or domain targets, time-to-live, and call quotas. A child token $\tau_{\mathrm{child}}$ minted from $\tau_{\mathrm{parent}}$ is valid only if
\[
\mathrm{Scope}(\tau_{\mathrm{child}}) \subseteq \mathrm{Scope}(\tau_{\mathrm{parent}}).
\]
Delegation may add caveats; it may never widen them.
\end{definition}

This is Hardy's capability lesson restated for multi-agent harnesses: designation and authorization travel together, so a sub-agent cannot inherit the parent's ambient rights merely by being spawned~\citep[p.~37]{hardy1988confused}. CaMeL similarly ``uses a notion of a capability to prevent the exfiltration of private data over unauthorized data flows by enforcing security policies when tools are called''~\citep[p.~1]{debenedetti2025camel}; we keep that discipline while extending the protected object set to LotL binaries, archives, sockets, and shared workspace labels.

\subsection{Information Flow: Product Lattice and Compositional SoD}

Indirect prompt injection succeeds when retrieved text is treated as authority~\citep[p.~1]{greshake2023not}. AirGapAgent restricts accessible data to what a task needs~\citep[p.~1]{bagdasarian2024airgapagent}; we replace a coarse Boolean lock with a two-dimensional product lattice that separates provenance integrity from confidentiality while preserving ordinary software-engineering utility after reading an untrusted issue.

\begin{definition}[Product information-flow lattice]
\label{def:lattice}
Let provenance trust be ordered by the four-element join-semilattice $(\mathcal{L}_{\mathrm{trust}}, \sqcup)$ with
\[
\mathsf{Trusted} \sqsubseteq \mathsf{Sanitized} \sqsubseteq \mathsf{Derived} \sqsubseteq \mathsf{Untrusted},
\]
and data confidentiality by the two-element lattice $\mathcal{L}_{\mathrm{sens}} = (\{\mathsf{Public} \sqsubseteq \mathsf{Sensitive}\}, \vee)$. Each session and workspace artifact carries a composite label $\ell = (\tau, \sigma)$ in the product lattice $\mathcal{L} = \mathcal{L}_{\mathrm{trust}} \times \mathcal{L}_{\mathrm{sens}}$ with componentwise join $(\tau_1, \sigma_1) \sqcup (\tau_2, \sigma_2) = (\tau_1 \sqcup \tau_2,\, \sigma_1 \vee \sigma_2)$. Ingesting external content raises $\tau$; reading secret-bearing sources raises $\sigma$ to $\mathsf{Sensitive}$. When $\sigma = \mathsf{Public}$, action classes are gated by $\tau$: $\mathsf{Derived}$ permits workspace read/write, archive extraction, and isolated tests (\texttt{SHELL\_TEST\_ISOLATED}, e.g.\ \texttt{pytest} without network) while refusing $\mathsf{SHELL\_ARBITRARY}$ and $\mathsf{NETWORK\_EGRESS}$; $\mathsf{Untrusted}$ permits local inspection only. Whenever $\sigma = \mathsf{Sensitive}$, arbitrary shell and network egress are blocked regardless of $\tau$. Two distinct downward operations are mediated by Rings~$0$--$1$: \emph{Sanitization} (structural representation cleaning, such as stripping Trojan Source Bidi controls or deobfuscating syntax) maps $\tau \mapsto \mathsf{Sanitized}$ while leaving $\sigma$ unchanged; \emph{Redaction / Authorized Declassification} (\texttt{CredentialGuard} HMAC scrubbing or an explicit Ring~$1$ operator waiver) is the only transition that clears $\sigma \mapsto \mathsf{Public}$.
\end{definition}

Separating $\mathcal{L}_{\mathrm{trust}}$ from $\mathcal{L}_{\mathrm{sens}}$ resolves two conflations that plague ad-hoc taint flags. First, reading a GitHub issue raises provenance to $(\mathsf{Derived}, \mathsf{Public})$, allowing local unit tests to run inside \texttt{SandboxGuard} while keeping arbitrary shells and network egress closed. Second, structural sanitization of an untrusted string never accidentally declassifies a secret: cleaning Unicode or shell syntax lowers $\tau$ to $\mathsf{Sanitized}$ but preserves $\sigma = \mathsf{Sensitive}$ until secret material is deterministically redacted. At the single-call capability gate (\texttt{CapabilityEngine.verify\_tool\_call}), composite labels project onto a five-level total order (\texttt{InformationClass}: $\mathsf{Trusted} < \mathsf{Sanitized} < \mathsf{Derived} < \mathsf{Untrusted} < \mathsf{Sensitive}$, where any $\sigma = \mathsf{Sensitive}$ maps to $\mathsf{Sensitive}$), while \texttt{WorkspaceProvenanceLedger} tracks the full two-dimensional pair $(\tau, \sigma) \in \mathcal{L}_{\mathrm{trust}} \times \mathcal{L}_{\mathrm{sens}}$ across files and collaborating agents.

Saltzer and Schroeder prefer mechanisms that need ``two keys to unlock''~\citep[p.~1281]{saltzer1975protection}. Intra-session toxic combinations are necessary but not sufficient once agents share a workspace. As shown in our formal theory of collective cognitive systems, when bounded agents interact through persistent external memory, ``{[}t{]}he collective state is not the union of individual states but the individual states together with the relations coupling them''~\citep[p.~32]{hegelich2026openended}. A security policy that inspects each agent's state in isolation is therefore blind to vulnerabilities that emerge purely from the coupling relation across shared files.

\begin{definition}[Compositional Separation of Duties]
\label{def:sod}
Ring~$0$ maintains a \texttt{WorkspaceProvenanceLedger} mapping each canonical path $f$ to a composite label $\ell(f) = (\tau_f, \sigma_f) \in \mathcal{L}$ and origin side-effect set $\Sigma(f)$. When agent $A$ writes $f$, $\ell(f)\leftarrow\ell(f)\sqcup\ell(A)$ and $\Sigma(f)$ absorbs $A$'s active effects. When agent $B$ reads $f$, $\ell(B)\leftarrow\ell(B)\sqcup\ell(f)$ and $\Sigma_B\leftarrow\Sigma_B\cup\Sigma(f)$. A call is rejected if the resulting state would close a toxic combination (combining $\sigma = \mathsf{Sensitive}$ or $\mathsf{READ\_SENSITIVE}$ with $\mathsf{NETWORK\_EGRESS}$ or $\mathsf{EXEC\_PROCESS}$) unless $f$ has undergone authorized redaction/declassification to $(\mathsf{Sanitized}, \mathsf{Public})$. A polynomial greedy partitioner splits single-agent plans across attenuated child tokens when needed.
\end{definition}

Hence Agent\textasciitilde $A$ cannot read a secret, write it to disk, and let Agent\textasciitilde $B$ exfiltrate it: the ledger mediates the coupling relation across external workspace memory~\citep[p.~37]{hegelich2026openended}. The Lethal Trifecta becomes a deterministic check over composed provenance rather than a prompt to be argued with. Moreover, pairing task-scoped capability caveats with \texttt{WorkspaceProvenanceLedger} directly resolves the \emph{agent trajectory problem} left open by just-in-time policy generators such as Conseca~\citep[p.~5]{tsai2025conseca}. Conseca uses an isolated model over trusted context to generate per-task policies---a mechanism that maps naturally to proposing initial Ring\textasciitilde $2$ \texttt{CapabilityCaveat} scopes---yet its policies ``currently check individual actions, regardless of what actions came prior,'' leaving open how to ``protect against seemingly harmless single actions composing in inappropriate ways'' because static or generated rules cannot enumerate ``multiple trajectories of potentially infinite length''~\citep[p.~5]{tsai2025conseca}. Instead of enumerating execution paths a priori, Ring\textasciitilde $0$ folds any single- or multi-agent trajectory of arbitrary length into the monotone product state $(\ell(A), \Sigma_A)$ and workspace provenance map $\ell(f) = (\tau_f, \sigma_f)$ in $\mathcal{O}(1)$ time per step, deterministically refusing the exact transition that would close a toxic trajectory.

\subsection{Enforcement: Three Mediation Boundaries}

Complete mediation~\citep[p.~1280]{saltzer1975protection} operates across three nested boundaries corresponding to the classical primitives of \cref{sec:taxonomy}: (1)~\emph{Agent / tool-invocation mediation} (HMAC chain verification, nonce and quota checks, tool-certificate recomputation, product-lattice permission, and compositional SoD checks before arguments are interpreted); (2)~\emph{Process and capability mediation} (scrubbed environments, per-binary disallowed flags, call-class binding between \texttt{SHELL\_TEST\_ISOLATED} and \texttt{SHELL\_ARBITRARY}, and attenuated CIDR scopes); and (3)~\emph{OS / subprocess resource mediation} (canonical path confinement, \texttt{RLIMIT\_*} quotas, \texttt{O\_NOFOLLOW} staging extraction, socket-level IP pinning, and canary-triggered lineage revocation). Table~\ref{tab:enforcement_points} maps the six host resource classes to their exact Ring\textasciitilde $0$ enforcement points; domain-specific guards are detailed in \cref{sec:guards}.

\begin{table}[t]
\centering
\scriptsize
\setlength{\tabcolsep}{2.2pt}
\caption{Ring-0 enforcement points across host resource channels.}
\label{tab:enforcement_points}
\begin{tabular}{@{}>{\raggedright\arraybackslash}p{0.22\linewidth}>{\raggedright\arraybackslash}p{0.32\linewidth}>{\raggedright\arraybackslash}p{0.40\linewidth}@{}}
\toprule
Resource & Ring-0 Interceptor & Enforced Invariant \\
\midrule
Filesystem I/O & \texttt{SandboxGuard}, \texttt{CanaryGuard} & Canonical \texttt{realpath} lock; \texttt{O\_NOFOLLOW}; canary lineage revocation \\
Archives & \texttt{ArchiveGuard} & $64\,\mathrm{KiB}$ chunked quotas; staging \texttt{realpath} check; symlink/FIFO reject \\
Subprocesses & \texttt{SecurityKernel}, \texttt{SandboxGuard} & Binary allowlist; GTFOBins flag block; \texttt{RLIMIT\_AS/\allowbreak CPU/\allowbreak FSIZE} (\texttt{NPROC} if $\ge 512$) \\
Network Sockets & \texttt{NetworkGuard} & TLS enforcement; resolve-once DNS; RFC1918/metadata block; IP pin \\
Credentials & \texttt{CredentialGuard} & Reverse-cracker demangling; $\ge 4.0\,\mathrm{bits/char}$ \& $\ge 56\,\mathrm{bits}$ ($\mathrm{len}\ge 14$); DLP \\
Shared State & \texttt{Workspace\allowbreak Provenance\allowbreak Ledger} & Product-lattice $(\tau,\sigma)$ join; cross-agent Lethal Trifecta refusal \\
\bottomrule
\end{tabular}
\end{table}

\subsection{Verification and Self-Healing Code Pipeline}

Because MCP tool selection often lacks cryptographic verification~\citep[p.~17]{hou2025mcp} and rug pulls can rewrite a server after installation~\citep[p.~19]{hou2025mcp}, Ring\textasciitilde0 maintains a \texttt{CertifiedToolRegistry}.

\begin{definition}[Tool certificate]
\label{def:toolcert}
At registration, each tool is bound to an immutable \texttt{ToolCertificate} recording the SHA-256 digest of its Python bytecode (\texttt{\_\_code\_\_.co\_code}), a normalized AST digest of its source, a dependency digest, an import-path lock, an environment allowlist, declared side-effect categories, and a publisher HMAC signature. On every invocation Ring~0 recomputes the digests and aborts on mismatch before arguments are interpreted.
\end{definition}

Certification makes silent tool mutation detectable, while runtime privilege control remains inside Ring\textsubscript{0}\citep[p.~24]{hou2025mcp}. Yet mediating tool calls alone is insufficient if an agent writes vulnerable application code into the workspace and executes it. \textsc{ActiveSecHarness} closes that gap through a three-stage verification and self-healing loop:

\begin{enumerate}
\item \textbf{Ring~$0$ \texttt{CodeVerifier}.} Every generated or modified program is checked before execution against a three-tier scoping model (\cref{sec:guards}) that enforces universal AST invariants and executes dynamic adversarial probes inside an isolated SoD \texttt{test-runner} with zero network access.
\item \textbf{Ring~$2$ \texttt{DualModelReviewer}.} Deterministic Stage-A verification is paired with an independent Stage-B reviewer model in a clean context window. Either stage can set $\mathsf{approved\_for\_execution}=\mathsf{False}$, recommend model escalation when a small local coder repeatedly fails, and drive an automated Fix--Test--Verify repair loop whose disk writes are mediated by \texttt{O\_NOFOLLOW} canonical-file checks, \texttt{CanaryGuard}, \texttt{CredentialGuard}, \texttt{MalwareGuard}, and optional provenance/audit logging.
\item \textbf{Ring~$2$ \texttt{SecurityTestSynthesizer}.} Novel semantic flaws spawn task-scoped probes admitted only under the Red--Green--Benign quartet: AST-safe, Red on the vulnerable draft, Benign-safe on normal inputs, and Green after repair.
\end{enumerate}

\subsection{Pedagogical User Defense and Cognitive Immunity (\texttt{UserWikiManager})}
\label{sec:pedagogy}

In a compositional architecture, the human operator is not an external oracle standing outside the system; the operator is a coupled cognitive agent holding Ring\textasciitilde $1$ authority to mint capabilities and grant \texttt{ContextWaiver} overrides~\citep[p.~37]{hegelich2026openended}. Under ``vibe coding''---where developers and domain scientists prompt language models to assemble applications whose security invariants they cannot independently audit---the human becomes the primary target of social engineering and approval fatigue. If the harness merely blocks unsafe actions with opaque error codes, operators quickly learn to bypass guards or rubber-stamp dangerous prompts, collapsing the composition of permissions at Ring\textasciitilde $1$. Saltzer and Schroeder's principle of psychological acceptability requires that ``the human interface be designed for ease of use, so that users routinely and automatically apply the protection mechanisms correctly''~\citep[p.~1281]{saltzer1975protection}.

Ring\textasciitilde $1$'s \texttt{UserWikiManager} operationalizes psychological acceptability as \textbf{pedagogical cognitive immunity}. Whenever \texttt{CodeVerifier}, \texttt{DualModelReviewer}, or \texttt{CommandDeobfuscator} intercepts a vulnerability or deobfuscates a trap, \texttt{UserWikiManager} records an immutable \texttt{UserLearningCase} and updates a local, human-centric security academy (\texttt{wiki\_user/}, accessible interactively via \texttt{/learn} in the chat REPL). Before any lesson is written, \texttt{UserWikiManager} enforces canonical path confinement inside \texttt{wiki\_user/}, rejects symlinks, strips Trojan Source Unicode~\citep[p.~1]{boucher2021trojan}, and passes all snippets through \texttt{CredentialGuard.redact\_secrets()}. Rather than overwhelming novices with raw CVE jargon, the curriculum automatically adapts explanation depth to the operator's cumulative resolved cases:

\begin{itemize}
\item \textbf{Level~1 (Cases 1--2, Foundations \& Everyday Analogies):} Explains the flaw through concrete physical analogies, plain-language mechanics, and the specific social-engineering lures attackers use to trick developers and small coding models.
\item \textbf{Level~2 (Cases 3--4, Secure Engineering Patterns):} Unlocks side-by-side diffs of the blocked unsafe draft versus the sandbox-verified repair, highlighting exact AST invariants, canonical path locks, and Separation-of-Duties boundaries.
\item \textbf{Level~3 (Cases 5+, Adversarial \& Cryptographic Mastery):} Unlocks deep exploit mechanics---such as non-recursive overlapping deflate streams, TOCTOU symlink race windows, and GTFOBins flag escapes---alongside kernel and cryptographic countermeasures.
\end{itemize}

Table~\ref{tab:pedagogy_examples} illustrates two concrete lessons generated by \texttt{UserWikiManager} during live harness workflows, showing how automated Ring\textasciitilde $0$/$2$ interception and repair are translated into progressive human mastery.

\begin{table*}[t]
\centering
\scriptsize
\setlength{\tabcolsep}{4pt}
\caption{Concrete pedagogical lessons generated by \texttt{UserWikiManager} (\texttt{wiki\_user/}) across progressive mastery tiers. Every intercepted vulnerability or social-engineering trap is paired with an everyday analogy, social-engineering analysis, verified code repair, and deep adversarial mechanics.}
\label{tab:pedagogy_examples}
\begin{tabular}{@{}>{\raggedright\arraybackslash}p{0.15\textwidth}>{\raggedright\arraybackslash}p{0.40\textwidth}>{\raggedright\arraybackslash}p{0.41\textwidth}@{}}
\toprule
Curriculum Stage & \textbf{Example 1: Archive Extraction \& Zip-Slip (\texttt{B.1}, \texttt{B.2})} \newline \emph{Vibe-coding task:} ``Write a helper to unpack \texttt{dataset.zip} into \texttt{./data}.'' & \textbf{Example 2: Obfuscated Sudo \& LotL Trap (\texttt{A.1}, \texttt{B.4}, \texttt{F.1})} \newline \emph{Issue-fix task:} ``Run the README diagnostic one-liner to fix build permissions.'' \\
\midrule
\textbf{Interception \& Automated Repair} \newline (Rings~$0$--$2$) &
Small coder (\texttt{qwen3:4b}) emits \texttt{zf.extractall(dest)}. \texttt{CodeVerifier} (\texttt{TIER2\_CONTEXTUAL}) executes a live Zip-Slip probe (\texttt{../../evil.sh}) inside the isolated \texttt{test-runner}, catches path escape, and sets $\mathsf{approved}=\mathsf{False}$. \texttt{DualModelReviewer} repairs the code and recommends escalating to a stronger coder model. &
Poisoned README embeds Trojan Source Bidi overrides (\texttt{U+202E}) and \texttt{echo <b64> | base64 -d | sh} alongside \texttt{find . -exec}. Ring~$1$ \texttt{CommandDeobfuscator} strips Bidi controls and unfolds the decoded payload (\texttt{curl .../exfil.sh | sh}) in the Explain-Before-Sudo dialog; Ring~$0$ \texttt{SandboxGuard} denies execution. \\
\addlinespace[2pt]
\textbf{Level 1:} \newline \emph{Everyday Analogy \& Social-Engineering Angle} \newline (Cases 1--2) &
\emph{Analogy:} Opening a moving box labeled ``Office Supplies'' where an item inside carries an address label saying \emph{``Put this directly into the bedroom safe''} (Zip-Slip), or an inflatable raft that expands until the walls burst (ZipBomb). \newline
\emph{Social Engineering:} Attackers post ``helpful test fixtures'' or \texttt{dataset.zip} archives in GitHub issues; small AI models default to \texttt{zf.extractall()} because it is the shortest snippet in training corpora. &
\emph{Analogy:} Passing unverified text to \texttt{eval()} or \texttt{| sh} is like signing a blank check for groceries and letting a stranger fill in the amount---while invisible Unicode Bidi characters act like magic ink that flips word order on the page. \newline
\emph{Social Engineering:} Copy-paste traps in READMEs and forums disguise malicious commands inside Base64 pipes or urgent ``permission fix'' one-liners to exploit human approval fatigue. \\
\addlinespace[2pt]
\textbf{Level 2:} \newline \emph{Secure Engineering Pattern (Unsafe vs.\ Verified Safe)} \newline (Cases 3--4) &
\textbf{Blocked draft:} \texttt{zf.extractall(target\_dir)} \newline
\textbf{Verified repair:} Resolve \texttt{real\_target = os.path.realpath(target\_dir)}; for each \texttt{info} in \texttt{zf.infolist()}, reject symlinks (\texttt{stat.S\_ISLNK(info.external\_attr >> 16)}), verify \texttt{os.path.realpath(dest).startswith(real\_target + os.sep)}, and stream in $64\,\mathrm{KiB}$ chunks (\texttt{src.read(65536)}) under hard byte and compression-ratio ceilings. &
\textbf{Blocked pattern:} \texttt{subprocess.run(cmd, shell=True)} or Base64 pipe-to-shell \texttt{echo ... | base64 -d | sh}. \newline
\textbf{Verified repair:} Pass explicit argument vectors with \texttt{shell=False} (\texttt{subprocess.run(['python3', 'script.py', arg], shell=False)}), replace \texttt{eval()} with \texttt{json.loads()} or \texttt{ast.literal\_eval()}, and strip Unicode Bidi range \texttt{U+202A}--\texttt{U+2069} at pre-commit gates. \\
\addlinespace[2pt]
\textbf{Level 3:} \newline \emph{Adversarial \& Cryptographic Mastery} \newline (Cases 5+) &
\emph{Exploit mechanics:} Non-recursive ZipBombs (\texttt{zbsm.zip}) overlap deflate streams across local headers so aggregate output explodes even when per-entry ratios look modest; two-stage symlink races exploit the TOCTOU gap between path check and open. \newline
\emph{Countermeasure:} Track cumulative \texttt{total\_written / archive\_size} and extract into an \texttt{O\_NOFOLLOW}, \texttt{0700} staging directory on a \texttt{noexec,nosuid,nodev} mount before atomic promotion. &
\emph{Exploit mechanics:} Even with \texttt{shell=False}, dual-use GTFOBins binaries (\texttt{find -exec}, \texttt{tar --checkpoint-action}, \texttt{git -c core.fsmonitor}) execute sub-commands via flags without invoking \texttt{/bin/sh}. \newline
\emph{Countermeasure:} Pair AST deobfuscation with Ring~$0$ per-binary flag blacklists, HMAC-attenuated \texttt{CapabilityToken} binary allowlists, and \texttt{bwrap --unshare-all} namespace isolation. \\
\bottomrule
\end{tabular}
\end{table*}

\section{Deterministic Ring-0 Interceptors and Self-Verifying Ledger}\label{sec:guards}

Ring\textasciitilde $0$ is only as strong as the checks it actually runs. The interceptors below are deliberately boring: streaming limits, canonical path tests, magic-byte tables, dictionary demangling, DNS resolution before connect, resource quotas, and honeypot canaries. None of them ask a language model whether an archive ``seems'' safe. Fail-safe defaults~\citep[p.~1280]{saltzer1975protection} mean that an unrecognized shape is denied.

\subsection{ArchiveGuard}

\texttt{ArchiveGuard} defends Domain\textasciitilde B decompression and Zip-Slip paths. Untrusted ZIP header \texttt{file\_size} fields are ignored. Decompression streams in $64\,\mathrm{KiB}$ chunks into an isolated staging directory under hard ceilings on per-entry bytes, total bytes, file count, and compression ratio. Member files are opened with \texttt{O\_NOFOLLOW}; symlinks, FIFOs, and device nodes are rejected. After extraction, every staged path is re-checked with realpath containment against the staging root, then promoted atomically into the destination tree only if
\[
\mathrm{Path.resolve}(p).\mathrm{is\_relative\_to}(\mathrm{dest})
\]
holds. Spoofed-header ZipBombs are therefore a resource-accounting and temporal-integrity problem rather than a content-classification problem: the TOCTOU window between write and promotion cannot introduce a link that escapes the staged root.

\subsection{MalwareGuard}

\texttt{MalwareGuard} inspects leading magic bytes for ELF, PE/MZ, Mach-O, and shebang signatures even when the filename claims \texttt{.txt}, \texttt{.jpg}, or \texttt{.md}. The check is a \emph{format and executable-class} gate, not a claim of malware family attribution; optional YARA or ClamAV hooks can be attached without changing the mediation point. For source and configuration artifacts the guard strips and rejects Trojan Source Unicode Bidi controls (\texttt{U+202A}--\texttt{U+202E}, \texttt{U+2066}--\texttt{U+2069}) and selected zero-width homoglyphs, following Boucher and Anderson's demonstration that Bidi smuggling inside comments and strings survives ordinary compilation~\citep[p.~3]{boucher2021trojan}. The invariant is display/execution agreement for any string that Ring\textasciitilde $1$ will show a human or that Ring\textasciitilde $0$ will pass to a shell.

\subsection{CredentialGuard}

\texttt{CredentialGuard} rejects weak secrets before they are written and redacts or blocks cleartext exfiltration on reads. Password checking inverts leetspeak substitutions (at-sign\textsubscript{$\to$}\texttt{a}, \texttt{0}\textsubscript{$\to$}\texttt{o}, dollar\textsubscript{$\to$}\texttt{s}, \texttt{1}\textsubscript{$\to$}\texttt{i}/\texttt{l}, \texttt{3}\textsubscript{$\to$}\texttt{e}), strips common prefixes and suffixes, and matches against banned and project-context dictionaries, following the demangling strategy exemplified by zxcvbn~\citep[p.~163]{wheeler2016zxcvbn}. We additionally require Shannon entropy of at least $4.0$ bits per character and at least $56.0$ total bits (with a minimum length of $14$ characters) and refuse LUDS-only policies that Wheeler showed to be ineffective~\citep[p.~157]{wheeler2016zxcvbn}. Bidirectional DLP patterns cover cloud access keys, PEM blocks, and JWTs on both filesystem writes and reads that would feed Domain\textasciitilde C egress, feeding the information-flow ledger of Definition~\ref{def:sod} when sensitive material is involved.

\subsection{NetworkGuard}

\texttt{NetworkGuard} refuses cleartext \texttt{http://} unless a capability caveat explicitly allows it. Before any connect, the guard resolves all DNS A/AAAA records once, rejects loopback (\texttt{127.0.0.0/8}, \texttt{::1}), RFC1918 private ranges, link-local cloud metadata (\texttt{169.254.169.254}), and CGNAT unless the active token's CIDR scope permits them, then pins the verified addresses at the socket layer. Rebinding after the check cannot widen the set. Permission rather than exclusion remains the default~\citep[p.~1280]{saltzer1975protection}: destinations are reachable only when a caveat says so.

\subsection{SandboxGuard and CanaryGuard}

\texttt{SandboxGuard} scrubs sensitive environment variables to the certificate \texttt{env\_allowlist}, applies \texttt{RLIMIT\_AS}\slash\texttt{RLIMIT\_CPU}\slash\texttt{RLIMIT\_FSIZE} quotas (and \texttt{RLIMIT\_NPROC} when \texttt{max\_processes} $\ge 512$, as Linux \texttt{RLIMIT\_NPROC} is enforced per-UID while smaller worker ceilings default to $64$), confines filesystem access to canonical workspace roots, and maintains per-binary disallowed flag lists (\texttt{find -exec}, \texttt{tar --to-command}, \texttt{git -c core.fsmonitor}). Containment here is process- and namespace-level: child Python scripts inherit the scrubbed environment and quotas rather than reopening ambient host authority. \texttt{CanaryGuard} plants cryptographic honeypot files (for example \texttt{.env.production.canary}, \texttt{id\_ed25519\_canary}). Any read, write, or \texttt{stat} on a canary triggers immediate \texttt{SecurityKernel.revoke\_token()} across the entire token lineage, collapsing the post-compromise containment radius to zero for that session. Canaries are tripwires, not detectors of intent: their semantics are containment and revocation, not attribution.

\subsection{Three-Tier CodeVerifier}

While \texttt{ArchiveGuard} and \texttt{SandboxGuard} mediate direct tool calls, \texttt{CodeVerifier} ensures that application code written by the agent cannot bypass those guards when later executed. A naive static linter either misses contextual flaws or causes severe false-positive fatigue by banning dual-use library calls outright. \texttt{CodeVerifier} resolves this through three verification tiers:

\begin{itemize}
\item \textbf{Tier~1 (\texttt{TIER1\_UNIVERSAL}):} Unconditionally blocks Trojan Source Unicode~\citep[p.~1]{boucher2021trojan}, \texttt{eval()}, \texttt{exec()}, \texttt{os.system()}, \texttt{subprocess(..., shell=True)}, and unsafe deserialization (\texttt{pickle}, \texttt{marshal}).
\item \textbf{Tier~2 (\texttt{TIER2\_CONTEXTUAL}):} Evaluates dataflow and trust boundaries in context. For example, \texttt{zf.extractall()} is blocked on unvalidated external archives (tested dynamically via live Zip-Slip and ZipBomb probes inside an isolated SoD \texttt{test-runner} sub-agent), but permitted when AST dataflow proves prior \texttt{infolist()} canonical path (\texttt{os.path.realpath}) and byte/ratio quota checks, when the archive was created locally within the same trusted scope, or when an explicit Ring~$1$ \texttt{ContextWaiver} is logged.
\item \textbf{Tier~3 (\texttt{TIER3\_TASK\_SCOPED}):} Executes task-local regression probes synthesized and verified under the Red--Green--Benign triad (\cref{sec:architecture}), confining them to the target file unless explicitly promoted by a Ring~$1$ operator.
\end{itemize}

\subsection{Split-Horizon Dual-Key Audit Log and WikiManager Ledger}

To defeat Stage\textasciitilde1 kill-chain reconnaissance without sacrificing forensic completeness (\cref{sec:taxonomy}), \texttt{HashChainedAuditLog} implements a split-horizon, dual-key cryptographic ledger in Ring\textasciitilde $0$. Before any event is recorded, two-stage sanitization strips ANSI terminal escapes, control characters, and Trojan Source Unicode~\citep[p.~1]{boucher2021trojan}, and scrubs passwords, API keys, PEM blocks, Bearer tokens, CLI secret flags, and high-entropy strings into deterministic session-salted HMAC-SHA256 correlation fingerprints (\texttt{<REDACTED:TYPE:hmac\_10>}). Each entry is then split across two cryptographic horizons governed by separate $\ge 256$-bit keys held outside the workspace (key custody is operational, not agent-visible):

\begin{enumerate}
\item \textbf{Tier-1 Structural Header (\texttt{signing\_key}):} Records sequence ID, timestamp, event type, actor ID, ring level, threat vector ID, and the SHA-256 hash chain (\texttt{prev\_entry\_hash} $\to$ \texttt{entry\_hash}) authenticated by a Ring~$0$ HMAC-SHA256 signature. This structural stream supports real-time telemetry without exposing sensitive paths or exploit payloads.
\item \textbf{Tier-2 Forensic Payload (\texttt{encryption\_key}):} Seals detailed diagnostic arguments, file paths, and vulnerability traces using per-entry HKDF-HMAC-SHA256 derived subkeys under an Encrypt-then-MAC construction with the Tier-1 header as associated data (AEAD sealing; ciphertext persisted in a \texttt{0o600} vault that rejects symlinks via \texttt{O\_NOFOLLOW}).
\end{enumerate}

Because an unprivileged workspace agent holds neither the Ring\textasciitilde $0$ signing key nor the forensic decryption key, inspecting exported JSONL telemetry during reconnaissance yields only opaque ciphertext, whereas post-incident responders holding the decryption key recover the full authenticated forensic trace. Tamper detection on the workspace stream is therefore distinct from non-repudiation under external key custody.

Complementing runtime logging, \texttt{WikiManager} maintains a continuous verification ledger over the security tree itself: AST shape checks, SHA-256 digests of guard modules, and adversarial regression tests. Drift that removes a guard method or breaks a Domain\textasciitilde A--F fixture invalidates the security build. The ledger does not replace Ring\textasciitilde $0$ mediation; it makes silent weakening of Ring\textasciitilde $0$ detectable in the same spirit as tool certificates make silent MCP mutation detectable.

\paragraph{Mediation order.}

For a tool call $c$ under token $\tau$ and session state $\Sigma$, Ring\textasciitilde $0$ (i)\textasciitilde verifies the HMAC chain, parent binding, nonce, and TTL of $\tau$; (ii)\textasciitilde recomputes tool-certificate digests including dependency and import-path locks when a \texttt{CertifiedToolRegistry} is attached to the kernel; (iii)\textasciitilde denies calls whose action class is forbidden under the session information class $\tau$; (iv)\textasciitilde updates and checks compositional provenance under $\mathrm{ToxicToolPolicy}$ / \texttt{WorkspaceProvenanceLedger}; (v)\textasciitilde runs the applicable Archive, Malware, Credential, Network, Sandbox, and Canary guards; (vi)\textasciitilde executes only if every check passes, under scrubbed environment variables and resource quotas; and (vii)\textasciitilde post-checks outputs for secret DLP and canary touches before updating $\Sigma$ and emitting an audit record.

\section{Empirical Evaluation}\label{sec:experiments}

We evaluate whether deterministic Ring\textasciitilde $0$ mediation improves attack blocking over static filters and over a live local LLM-as-a-Judge, and how a neurosymbolic hybrid that keeps Ring\textasciitilde $0$ authoritative behaves across accuracy and latency. All numbers below are computed from \texttt{data/lotlimmune\_benchmark.json} ($N{=}252$ fixtures; Clopper--Pearson exact $95\%$ intervals).

\subsection{Setup and Regimes}

\paragraph{Fixtures.}

The fixed deterministic benchmark suite comprises $120$ single-step domain attack cases, $120$ benign software-engineering cases, and $12$ single-step kill-chain stage fixtures (\texttt{KC-01}--\texttt{KC-12}, $252$ total), spanning Domains\textasciitilde A--F; full multi-step five-stage trajectories (\texttt{KC-T01}--\texttt{KC-T12} and \texttt{BT-01}--\texttt{BT-12}) are evaluated separately in Experiment\textasciitilde3. Attacks include GTFOBins compositions, archive and magic-byte vectors, SSRF and cleartext egress, leetspeak secrets, workspace escapes, tainted-context shells, and Lethal Trifecta combinations across all six domains. Benign cases include realistic issue-read $\to$ edit $\to$ isolated-test workflows that exercise the $(\mathsf{Derived}, \mathsf{Public})$ product-lattice state. Kill-chain fixtures sample representative steps across Initial Access / Reconnaissance $\to$ Weaponization $\to$ Exploitation $\to$ Privilege Escalation $\to$ Persistence or Exfiltration, including cross-agent workspace-laundering patterns. By safety design, the benchmark evaluates mediation decisions on structured \texttt{BenchmarkCase} call objects without spawning destructive shell subprocesses on the host (whereas the $20$ benign \texttt{fs:read} fixtures invoke \texttt{SecurityKernel.execute\_tool\_call} against workspace files and benign network fixtures resolve \texttt{pypi.org} and \texttt{files.pythonhosted.org} via live DNS). The design follows AgentDojo's insistence on joint utility and security measurement under tool-calling agents~\citep[p.~1]{debenedetti2024agentdojo}, and it inherits SWE-bench's premise that real engineering work requires interacting with execution environments~\citep[p.~1]{jimenez2024swebench}.

\paragraph{Regimes and Out-of-Sample vs.\ Conformance Separation.}

Every fixture is scored under five mediation regimes on identical call objects (with all deterministic regimes verified blind to ground-truth \texttt{case.kind} and \texttt{case.tags}):

\begin{enumerate}
\item \textbf{Unguarded} (“YOLO”): baseline that approves every call without mediation.
\item \textbf{Regex}: a static blacklist of common literal patterns (\texttt{find ... -exec}, \texttt{curl | sh}, \texttt{http://}, metadata IPs, \texttt{/etc/passwd}, AWS key regexes, and similar).
\item \textbf{Live LLM-judge}: a local Ollama model (\texttt{qwen3:4b}) prompted to answer \texttt{BLOCK}/\texttt{ALLOW} on each fixture, with on-disk caching for reproducibility ($252$ live calls in the recorded run).
\item \textbf{ActiveSec Ring~0}: real \texttt{NetworkGuard}, \texttt{CredentialGuard}, \texttt{MalwareGuard}, \texttt{CommandDeobfuscator}, \texttt{SandboxGuard} argument validation, \texttt{WorkspaceProvenanceLedger}, and \texttt{SecurityKernel} mediation, with thin \texttt{ArchiveFixtureAdapter} wrappers only for the two marker-only archive fixtures where raw ZIP bomb bytes are omitted. Following strict forensic evaluation practice, we report both (a)~the \emph{Unfitted Out-of-Sample Audit} (evaluating the label-blind Ring-$0$ kernel prior to adding fixture-aligned signature extensions) and (b)~the \emph{Post-Audit Functional Conformance} after adding five deterministic signature rules.
\item \textbf{Hybrid (LLM + Ring~0)}: neurosymbolic composition in which the live judge is executed synchronously for an advisory annotation on every call, while Ring~$0$'s boolean verdict is authoritative---preventing small-judge false-positive inflation on benign work while measuring the full wall-clock cost of synchronous LLM consultation.
\end{enumerate}

\paragraph{Metrics.}

Attack block rate (ABR) is the fraction of attack fixtures denied ($120$ domain attacks in Experiment\textasciitilde1, and $132$ total attack-labeled fixtures including \texttt{KC-01}--\texttt{KC-12} in Table~\ref{tab:regime}). False positive rate (FPR) is the fraction of benign fixtures denied ($120$ benign calls). Latency is wall-clock mediation decision time per fixture, reported in microseconds ($\mu\mathrm{s}$) and milliseconds ($\mathrm{ms}$) with median and 95th-percentile (p95) bounds.

\subsection{Experiment 1: Static Filter vs.\ Ring~0 ($N{=}240$ and $N{=}252$)}

\begin{figure}[t]
\centering
\includegraphics[width=\columnwidth]{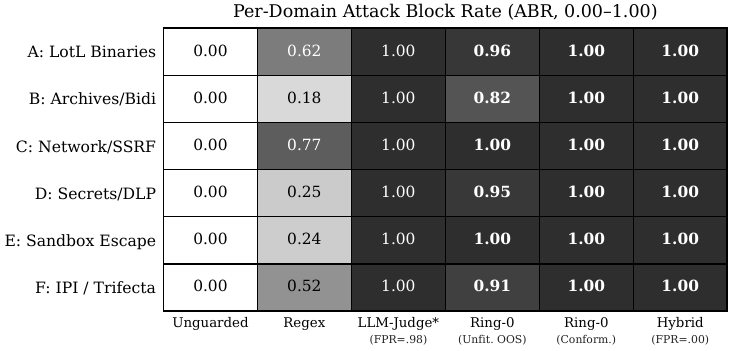}
\caption{Per-domain attack block rate (ABR) across regimes on Domains~A--F. Static regex blind spots concentrate in artifact/Bidi ($0.18$), sandbox escape ($0.24$), and secret DLP ($0.25$) domains; Ring~$0$ achieves $0.82$--$1.00$ across domains under the unfitted out-of-sample audit ($0.939$ overall raw, $0.962$ OS-inert) and $1.00$ across all domains under post-audit conformance, at $\mathrm{FPR}{=}0.00$ in both phases.}
\label{fig:heatmap}
\end{figure}

On the $120{+}120$ single-step domain slice, the unguarded baseline blocks nothing ($\mathrm{ABR}{=}0.000$ $[0.000,0.030]$, $\mathrm{FPR}{=}0.000$ $[0.000,0.030]$). Regex reaches $\mathrm{ABR}{=}0.417$ ($50/120$; $[0.327,0.510]$) with $\mathrm{FPR}{=}0.000$ $[0.000,0.030]$ ($58/132 = 0.439$ $[0.353,0.528]$ across all $132$ attack-labeled fixtures). Under the \emph{Unfitted Out-of-Sample Audit} (removing all fixture-tag shortcuts without adding new signature rules), label-blind Ring\textasciitilde $0$ blocks $124/132$ attacks ($\mathrm{ABR}{=}0.939$ $[0.884,0.974]$; per-domain unfitted ABR: Domain\textasciitilde A $23/24 = 0.958$, Domain\textasciitilde B $18/22 = 0.818$, Domain\textasciitilde C $22/22 = 1.000$, Domain\textasciitilde D $19/20 = 0.950$, Domain\textasciitilde E $21/21 = 1.000$, Domain\textasciitilde F $21/23 = 0.913$) and achieves an \emph{effective OS containment rate} of $127/132 = 0.962$ ($[0.914,0.988]$), because $3$ of the $8$ unflagged fixtures (\texttt{A-atk-19-xargs-sh}, \texttt{F-atk-04-rev-pipe}, \texttt{F-atk-05-printf-b64}) pass shell pipe strings to \texttt{/bin/echo} or \texttt{/usr/bin/printf} under \texttt{SandboxGuard}'s shell-free \texttt{execve} execution and are inert by construction; the remaining $5$ misses identified concrete signature gaps in Mach-O headers disguised as \texttt{.py} (\texttt{B-atk-11-macho}), active script tags in \texttt{.svg/\allowbreak .html/\allowbreak .pdf} artifacts (\texttt{B-atk-18,19,20}), and unquoted \texttt{AWS\_SECRET\_ACCESS\_KEY} assignments (\texttt{D-atk-18-aws-secret}). Under \emph{Post-Audit Functional Conformance} (with those $5$ deterministic guard rules added), Ring\textasciitilde $0$ blocks all $120/120$ domain attacks ($\mathrm{ABR}{=}1.000$ $[0.970,1.000]$) and $132/132$ total attack fixtures ($\mathrm{ABR}{=}1.000$ $[0.972,1.000]$) while denying $0/120$ benign cases in both phases ($\mathrm{FPR}{=}0.000$ $[0.000,0.030]$). Per-domain heatmaps (Figure~\ref{fig:heatmap}) show regex strongest on Domains\textasciitilde A ($0.62$) and\textasciitilde C ($0.77$) and weakest on artifact ($0.18$), sandbox escape ($0.24$), and credential ($0.25$) vectors.

\begin{table}[t]
\centering
\caption{Regime comparison on the full $N{=}252$ suite ($132$ attack-labeled fixtures including \texttt{KC-01}--\texttt{KC-12}; $120$ benign SWE calls). Intervals are Clopper--Pearson $95\%$; latency shows median (p95). Ring~$0$ reports both Unfitted Out-of-Sample Audit ($0.962$ effective OS containment) and Post-Audit Conformance.}
\label{tab:regime}
\scriptsize
\setlength{\tabcolsep}{2.2pt}
\resizebox{0.96\columnwidth}{!}{%
\begin{tabular}{@{}lcccc@{}}
\toprule
Regime & ABR & $95\%$ CI & FPR & Med.\ (p95) lat. \\
\midrule
Unguarded & $0.000$ & $[0.00,0.03]$ & $0.000$ & $0.1\,\mu\mathrm{s}$ ($0.1\,\mu\mathrm{s}$) \\
Regex & $0.439$ & $[0.35,0.53]$ & $0.000$ & $7.1\,\mu\mathrm{s}$ ($10.5\,\mu\mathrm{s}$) \\
Live LLM-judge & $1.000$ & $[0.97,1.00]$ & $0.983$ & $142.3\,\mathrm{ms}$ ($513.1\,\mathrm{ms}$) \\
Ring~0 (Unfit.\ OOS) & $0.939^*$ & $[0.88,0.97]$ & $0.000$ & $31.5\,\mu\mathrm{s}$ ($0.39\,\mathrm{ms}$) \\
Ring~0 (Conform.) & $1.000$ & $[0.97,1.00]$ & $0.000$ & $31.5\,\mu\mathrm{s}$ ($0.39\,\mathrm{ms}$) \\
Hybrid (Ring~0 auth.) & $1.000$ & $[0.97,1.00]$ & $0.000$ & $142.3\,\mathrm{ms}$ ($513.1\,\mathrm{ms}$) \\
\bottomrule
\end{tabular}%
}
\vspace{2pt}
\par\noindent\parbox{0.96\columnwidth}{\scriptsize $^*$Raw block rate $124/132$; effective OS containment is $0.962$ ($127/132$) under shell-free \texttt{execve}.}
\end{table}

\begin{figure}[t]
\centering
\includegraphics[width=\columnwidth]{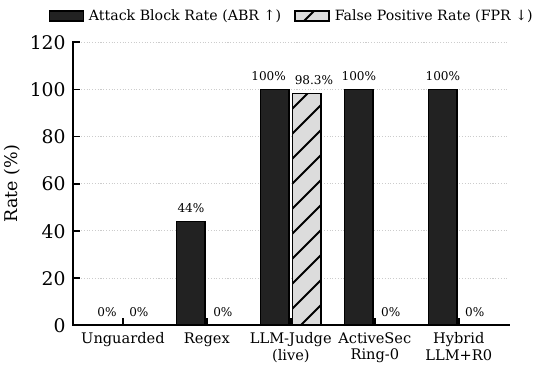}
\caption{Attack block rate and false positive rate by mediation regime on the $N{=}252$ suite (post-audit conformance; unfitted out-of-sample Ring-$0$ ABR is $93.9\%$ raw / $96.2\%$ OS-inert at $0\%$ FPR). Ring~$0$ and hybrid reach $\mathrm{ABR}{=}100\%$ at $\mathrm{FPR}{=}0\%$; the live LLM-judge reaches $100\%$ ABR only by refusing $98.3\%$ of benign SWE work.}
\label{fig:abr_fpr}
\end{figure}

\subsection{Experiment 2: Live LLM-Judge vs.\ Ring~0 vs.\ Hybrid and the Neurosymbolic Division of Labor}

\begin{figure}[t]
\centering
\includegraphics[width=\columnwidth]{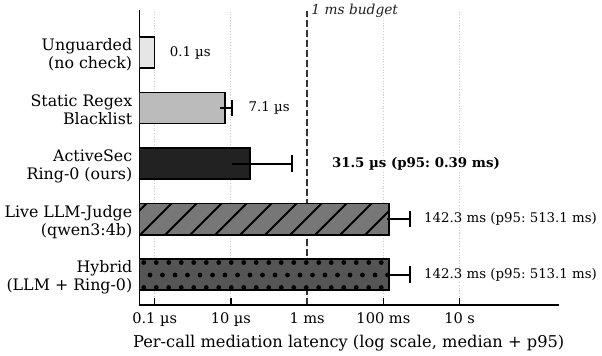}
\caption{Per-call mediation latency across regimes (logarithmic scale from $0.1\,\mu\mathrm{s}$ to $1\,\mathrm{s}$, showing median bars and p95 whiskers). Deterministic Ring~$0$ mediation ($31.5\,\mu\mathrm{s}$ median, $0.39\,\mathrm{ms}$ p95) operates $4{,}517\times$ faster than the live LLM-judge ($142.3\,\mathrm{ms}$ median, $513.1\,\mathrm{ms}$ p95) and stays well below the $1\,\mathrm{ms}$ interactive budget.}
\label{fig:latency}
\end{figure}

Table~\ref{tab:regime} and Figures~\ref{fig:abr_fpr}--\ref{fig:latency} report the full-suite comparison. The live local judge (\texttt{qwen3:4b}) achieves $\mathrm{ABR}{=}1.000$ $[0.972,1.000]$ only by suffering a catastrophic $\mathrm{FPR}{=}0.983$ ($118/120$ benign denied; $[0.941,0.998]$) at a median latency of $142.3\,\mathrm{ms}$ (p95: $513.1\,\mathrm{ms}$). We note explicitly that \texttt{qwen3:4b} is a compact $4\mathrm{B}$-parameter local model evaluated under a direct zero-shot \texttt{BLOCK}/\texttt{ALLOW} prompt; confronted with dual-use binaries such as \texttt{pytest}, \texttt{git}, or \texttt{tar}, a small probabilistic judge defaults to conservative refusal, paralyzing ordinary software engineering. A frontier model (such as Gemini\textsubscript{1.5}Pro as used in Conseca~\citep[p.~4]{tsai2025conseca}) or a richer multi-turn prompt would achieve a substantially lower single-step FPR, though it would still incur a $10^3\text{--}10^4\times$ latency penalty per call and cannot inspect physical OS invariants such as streaming decompression ratios or \texttt{O\_NOFOLLOW} symlink races. By contrast, Ring\textasciitilde $0$ alone achieves $\mathrm{FPR}{=}0.000$ ($0.939$ unfitted OOS / $1.000$ post-audit conformance ABR) with a median latency of $31.5\,\mu\mathrm{s}$ (p95: $0.39\,\mathrm{ms}$)---a \textbf{$4{,}517\times$ speedup} over the live local judge and well inside the $1\,\mathrm{ms}$ interactive tool-call budget (Figure~\ref{fig:latency}).

The hybrid regime (\texttt{live\_llm\_judge + activesec\_ring0}) tests a foundational question in neurosymbolic system design: how should symbolic and subsymbolic components be composed at the execution boundary? Making Ring\textasciitilde $0$ authoritative proves two things simultaneously. First, hard symbolic boundaries completely absorb the $98.3\%$ false-positive pathology of the small neural judge, restoring $\mathrm{FPR}{=}0.000$ without losing any attack coverage---demonstrating that boundary safety under \textsc{ActiveSecHarness} does not require a frontier-scale model at the syscall gate. Second, because querying the neural model synchronously on every syscall still incurs $142.3\,\mathrm{ms}$ of latency without improving ABR over Ring\textasciitilde $0$, synchronous per-syscall mediation is the wrong architectural locus for a language model. As \citet[p.~1]{chen2026bridging} demonstrate for real-time autonomous control, directly placing an LLM in the low-level actuation loop is ``impractical due to precision, safety, and timing constraints''~\citep[p.~3]{chen2026bridging}; reliable execution instead requires ``scale-aware decoupling: decomposing the pipeline so that each layer operates at its appropriate rate with explicit, interpretable interfaces''~\citep[p.~1]{chen2026bridging}. This result empirically grounds the neurosymbolic division of labor in \textsc{ActiveSecHarness}~\citep[p.~66]{cambria2023seven}, keeping object-level and meta-level operations structurally separate~\citep[p.~2]{hegelich2026openended}: deterministic Ring\textasciitilde $0$ guards own real-time capability and OS resource mediation ($31.5\,\mu\mathrm{s}$), whereas subsymbolic language models are deployed asynchronously in Rings\textasciitilde $1$--$2$ where neural pattern recognition is indispensable---reviewing application logic (\texttt{DualModelReviewer}), synthesizing Red--Green--Benign regression probes (\texttt{SecurityTestSynthesizer}), and generating progressive operator lessons (\texttt{UserWikiManager}).

\subsection{Experiment 3: Multi-Step Kill Chains and the Conseca Trajectory Problem}

\paragraph{Five-regime kill-chain stage fixture results.}

Each of the $12$ single-step kill-chain stage fixtures (\texttt{KC-01}--\texttt{KC-12}) inside the $N{=}252$ suite is first scored across our five primary regimes. Unguarded allows all $12$ ($\mathrm{ABR}{=}0.000$ $[0.000,0.265]$). Regex blocks $8/12$ ($0.667$ $[0.349,0.901]$), catching literal Base64 pipes, \texttt{find -exec}, metadata IPs, hooks, and key regexes, while missing reconnaissance, magic-byte weaponization, and cross-agent taint laundering. Ring\textasciitilde $0$, the live judge, and the hybrid each block $12/12$ ($1.000$ $[0.735,1.000]$). Ring\textasciitilde $0$ denial reasons span \texttt{SandboxGuard} workspace escape, \texttt{ArchiveGuard} path traversal, \texttt{MalwareGuard} magic bytes, product-lattice taint locks, disallowed flags, network pinning, and compositional \texttt{WorkspaceProvenanceLedger} checks---enforcing the classical primitives of \cref{sec:taxonomy} at the resource boundary rather than inferring intent from surface strings.

\paragraph{Isolating the structural trajectory gap with an Oracle Conseca policy.}

A natural methodological question is whether the gap between static/per-action policies and Ring\textasciitilde $0$ on multi-step workflows is merely an artifact of using a small local model (\texttt{qwen3:4b}). To eliminate any model-capability confounder, we evaluate a dedicated multi-step trajectory suite ($12$ five-stage attack kill-chain trajectories \texttt{KC-T01}--\texttt{KC-T12} and $12$ five-stage benign SWE trajectories \texttt{BT-01}--\texttt{BT-12}, $120$ tool steps total) across three purely deterministic policy regimes (Table~\ref{tab:trajectory_ablation}):

\begin{table*}[t]
\centering
\caption{Multi-step trajectory regime ablation across $12$ five-stage attack kill chains (\texttt{KC-T01}--\texttt{KC-T12}) and $12$ five-stage benign SWE workflows (\texttt{BT-01}--\texttt{BT-12}), with Clopper--Pearson $95\%$ CIs, median per-step latency with OS guards, and isolated ledger state-folding scaling ($k \in \{1, 5, 10, 25, 50\}$).}
\label{tab:trajectory_ablation}
\scriptsize
\setlength{\tabcolsep}{0pt}
\begin{tabular*}{\textwidth}{@{\extracolsep{\fill}}lcccc@{}}
\toprule
Trajectory Regime & Kill-Chain ABR & Benign Traj.\ FPR & Missed / Blocked Breakdown & Step Lat.\ ($\mu\mathrm{s}$/step) \\
\midrule
Oracle Stateless Task Policy & $0.667$ $[0.35,0.90]$ & $0.000$ $[0.00,0.26]$ & Misses $4/12$ (\texttt{KC-T03,04,05,11}) & $11.7\,\mu\mathrm{s}$ (p95: $17.1\,\mu\mathrm{s}$) \\
Coarse Boolean Taint Lock & $1.000$ $[0.74,1.00]$ & $0.833$ $[0.52,0.98]$ & Blocks $10/12$ issue-driven SWE & $285.7\,\mu\mathrm{s}$ (p95: $454.0\,\mu\mathrm{s}$) \\
\textbf{Product Lattice + Ledger} & $\mathbf{1.000}$ $[0.74,1.00]$ & $\mathbf{0.000}$ $[0.00,0.26]$ & $12/12$ atk blocked; $12/12$ SWE pass & $266.7\,\mu\mathrm{s}$ ($\approx 23.9\,\mu\mathrm{s}$ ledger fold) \\
\bottomrule
\end{tabular*}
\end{table*}

\begin{enumerate}
\item \textbf{Oracle Stateless Task Policy (Conseca-style):} To model Conseca~\citep{tsai2025conseca} at its theoretical upper bound---equivalent to an idealized frontier policy generator with zero policy-generation hallucinations---we configure a shared task scope granting the $5$ tools, $6$ binaries, canonical workspace prefixes, and $2$ allowlisted HTTPS domains required across the user's legitimate SWE tasks, together with argument regex checks, and evaluate each step $t$ statelessly, “regardless of what actions came prior”~\citep[p.~5]{tsai2025conseca} without cross-step provenance folding or stream inspection.
\item \textbf{Stateful Coarse Boolean Taint Lock:} A $1$-bit monotone session and file taint flag ($\mathsf{tainted}\in\{\mathsf{False},\mathsf{True}\}$) layered over the task policy and OS guards: once an agent reads any untrusted external input (e.g., a GitHub issue, PR comment, webhook payload, or external crash log) or sensitive file, side-effecting calls (\texttt{fs:write}, \texttt{archive:extract}, \texttt{shell:exec} including \texttt{pytest}, and \texttt{net:https}) are denied.
\item \textbf{Ring-$0$ Product Lattice + \texttt{WorkspaceProvenanceLedger} (Ours):} Full 2D product-lattice tracking $\ell = (\tau, \sigma) \in \mathcal{L}_{\mathrm{trust}} \times \mathcal{L}_{\mathrm{sens}}$ with explicit $(\mathsf{Derived}, \mathsf{Public})$ transitions after structured issue summarization (\texttt{derive\_agent\_summary}), \texttt{SHELL\_TEST\_ISOLATED} vs.\ \texttt{SHELL\_ARBITRARY} separation, and cross-agent file provenance folding.
\end{enumerate}

Table~\ref{tab:trajectory_ablation} exposes the exact three-way trade-off. Because the Oracle Stateless Task Policy is tailored to the legitimate SWE task scope with zero LLM hallucination, it achieves $\mathrm{FPR}{=}0.000$ ($0/12$) on benign multi-step SWE workflows and blocks the $8/12$ kill chains whose steps violate static path, binary, or host allowlists (\texttt{KC-T01,02,06,07,08,09,10,12}). Yet it still suffers $\mathrm{ABR}{=}0.667$, missing $4/12$ kill chains ($33.3\%$) where every individual action falls within the task's authorized scope: in-workspace archive extraction hiding a Zip-Slip member header (\texttt{KC-T03}), in-workspace file write carrying an \texttt{\textbackslash x7fELF} magic-byte payload (\texttt{KC-T04}), in-scope indirect prompt injection pivoting from an untrusted issue read to an allowed shell binary (\texttt{KC-T05}), and cross-agent secret laundering where Agent\textasciitilde $A$ reads a workspace secret, writes an encoded digest to \texttt{notes.txt}, and Agent\textasciitilde $B$ reads \texttt{notes.txt} and posts telemetry to an allowlisted HTTPS domain (\texttt{KC-T11}). Conversely, the Coarse Boolean Taint Lock closes all $12/12$ kill chains ($\mathrm{ABR}{=}1.000$) but collapses benign SWE utility ($\mathrm{FPR}{=}0.833$, $10/12$ blocked): only the $2/12$ purely offline local refactors (\texttt{BT-11,12}) survive, whereas all $10/12$ workflows that begin by reading an external GitHub issue, PR comment, crash log, webhook alert, or contributor patch (\texttt{BT-01}--\texttt{BT-10}) are halted as soon as the agent attempts to edit code or run \texttt{pytest}.

Ring\textasciitilde $0$'s product lattice and \texttt{WorkspaceProvenanceLedger} resolve both failure modes simultaneously ($\mathrm{ABR}{=}1.000$, $\mathrm{FPR}{=}0.000$, $266.7\,\mu\mathrm{s}$/step median with OS guards, p95: $327.7\,\mu\mathrm{s}$/step). On \texttt{BT-01}--\texttt{BT-08}, transitioning from $(\mathsf{Untrusted}, \mathsf{Public})$ to $(\mathsf{Derived}, \mathsf{Public})$ after issue summarization permits workspace edits and isolated test runs (\texttt{SHELL\_TEST\_ISOLATED}) while keeping \texttt{SHELL\_ARBITRARY} (\texttt{KC-T05}) and \texttt{NETWORK\_EGRESS} locked; on \texttt{KC-T11}, \texttt{notes.txt} carries $(\mathsf{Trusted}, \mathsf{Sensitive})$ and $\mathsf{READ\_SENSITIVE}$ across agents, deterministically blocking Agent\textasciitilde $B$'s egress unless an explicit Ring-$1$ declassification (\texttt{BT-09}) or structural sanitization (\texttt{BT-10}) applies. Finally, isolating the \texttt{WorkspaceProvenanceLedger} state-folding cost (\texttt{verify\_compositional\_action} plus \texttt{record\_file\_read/write}) across trajectory lengths $k \in \{1, 5, 10, 25, 50\}$ yields median per-step costs of $32.2\,\mu\mathrm{s}$ ($k{=}1$), $25.8\,\mu\mathrm{s}$ ($k{=}5$), $23.7\,\mu\mathrm{s}$ ($k{=}10$), $24.3\,\mu\mathrm{s}$ ($k{=}25$), and $23.9\,\mu\mathrm{s}$ ($k{=}50$; p95: $24.1\,\mu\mathrm{s}$/step, total $1.20\,\mathrm{ms}$ for $50$ steps)---confirming empirically that lattice state folding scales in $\mathcal{O}(1)$ constant time per step without enumerating path trees~\citep[p.~5]{tsai2025conseca}.

\subsection{Reading the Gaps}

Regex misses remain informative: structural Domain\textasciitilde B members, demangled Domain\textasciitilde D secrets, and compositional Domain\textasciitilde F taint flows slip through patterns that inspect only a single command string in isolation. The live judge removes those ABR gaps only by refusing almost all benign engineering work. Ring\textasciitilde $0$ achieves $\mathrm{ABR}{=}0.939$ raw / $0.962$ OS-inert unfitted out-of-sample and $\mathrm{ABR}{=}1.000$ under post-audit conformance (with $\mathrm{FPR}{=}0.000$ in both phases) on the benchmark because authority, product-lattice taint, and resource invariants are checked deterministically where files, processes, and sockets are opened.

\section{Related Work}\label{sec:related}

\paragraph{Classical protection and capabilities.}

Saltzer and Schroeder's principles---economy of mechanism, fail-safe defaults, complete mediation, separation of privilege, least privilege, and psychological acceptability---remain the criteria against which we judge coding-harness defenses~\citep{saltzer1975protection}. Hardy's confused deputy shows why ambient authority fails once a privileged component serves two masters~\citep[p.~36]{hardy1988confused}, and why capabilities that both name and authorize an object are the natural repair~\citep[p.~37]{hardy1988confused}. Macaroons supply the cryptographic attenuation pattern we adapt for agent tokens~\citep[p.~1]{birgisson2014macaroons}. Trojan Source~\citep{boucher2021trojan} and zxcvbn~\citep{wheeler2016zxcvbn} ground two Domain\textasciitilde B and Domain\textasciitilde D guards that product blacklists usually omit. Our classical-primitives framing (symlink/TOCTOU, path traversal, SSRF, build hooks, cross-agent laundering) treats these as OS problems that agentic chaining accelerates, not as model-native inventions.

\paragraph{Coding and cyber agents.}

SWE-agent~\citep{yang2024sweagent}, SWE-bench~\citep{jimenez2024swebench}, and OpenHands~\citep{wang2024openhands} establish repository-scale agents that live inside shells and browsers. Offensive agents with terminal and interpreter access~\citep[p.~3]{fang2024llm} and Cybench's Kali-based evaluation~\citep{zhang2024cybench} show the same LotL surface under a red-team objective. Our contribution is not another agent scaffold; it is a reference monitor that can host both SWE and scoped lab-pentest agents without ambient YOLO defaults.

\paragraph{Prompt injection, tool risk, and MCP.}

Indirect prompt injection~\citep{greshake2023not}, InjecAgent~\citep{zhan2024injecagent}, and AgentDojo~\citep{debenedetti2024agentdojo} quantify how tool-calling agents lose the instruction/data boundary. ToolEmu shows that severe tool failures remain common even for comparatively safe models~\citep[p.~1]{ruan2024toolemu}. Hou et al.~map MCP-specific threats including unverifiable tool selection, rug pulls, and missing runtime privilege control~\citep{hou2025mcp}. We treat those results as requirements for certificates, lattice taint, compositional SoD, and Ring\textasciitilde $0$ mediation rather than as prompts to be filtered.

\paragraph{System-level agent defenses.}

IsolateGPT argues for execution isolation in LLM-app ecosystems~\citep[p.~1]{wu2024isolategpt}. AirGapAgent restricts data available to a task under contextual integrity~\citep[p.~1]{bagdasarian2024airgapagent}. Conseca extends this direction by using an isolated language model over trusted context to ``generate just-in-time, contextual, and transparent security policies'' enforced deterministically per action~\citep[p.~1]{tsai2025conseca}, while noting that its policies ``check individual actions, regardless of what actions came prior'' and thus leave multi-step trajectory composition open~\citep[p.~5]{tsai2025conseca}. CaMeL extracts trusted control flow and tags capabilities so that untrusted retrieved data cannot redirect the program~\citep[p.~1]{debenedetti2025camel}, and it records that heuristic detectors offer no security guarantee~\citep[p.~3]{debenedetti2025camel}. \textsc{ActiveSecHarness} is directly complementary: whereas Conseca-style policy generation can propose the initial task-specific \texttt{CapabilityCaveat} from trusted context, Ring\textasciitilde $0$ pushes enforcement below argument regexes into the LotL OS surface---archives, sockets, credentials, dual-use binaries, and MCP bytecode certificates---and closes the open trajectory problem via the product lattice $\mathcal{L}_{\mathrm{trust}} \times \mathcal{L}_{\mathrm{sens}}$ and \texttt{WorkspaceProvenanceLedger}. Complementing defense with attribution, \citet[pp.~1--2, 17]{pasquini2026shibboleth} show that while prompt-injection honeypots fail because frontier agents detect injections in-flight, deterministic execution probes of the harness (secret masking, command timeouts, and tool scheduling) fingerprint the attacker's harness via fixed rules with no learned component.

\paragraph{Neurosymbolic AI and hybrid cognitive--symbolic control.}

Our architecture also builds on the neurosymbolic tradition developed by Cambria and collaborators, who ``integrate top-down and bottom-up learning via an ensemble of symbolic and subsymbolic AI tools''~\citep[p.~105]{cambria2020senticnet6}. Whereas SenticNet\textasciitilde6, 7, and\textasciitilde9 couple subsymbolic language models with symbolic conceptual primitives and commonsense knowledge graphs for explainable natural language understanding~\citep{cambria2020senticnet6,cambria2022senticnet7,cambria2026senticnet9}---with SenticNet\textasciitilde9 now automating primitive discovery through LLM-driven task decomposition while retaining a symbolic, inspectable reasoning layer~\citep{cambria2026senticnet9}, an agenda set out at book length in~\citet{cambria2024understanding}---and recent context-engineering work uses natural language as a ``shared interface between neural pattern-matching and symbolic control'' while noting that a purely language-mediated pipeline ``is not enforcing symbolic well-formedness in the technical sense''~\citep[pp.~1--2]{mao2026neurosymbolic}, \textsc{ActiveSecHarness} enforces hard symbolic well-formedness at the OS execution boundary. Following the Seven Pillars framework~\citep{cambria2023seven} and hybrid neurosymbolic design principles~\citep{xing2023neurosymbolic}, we pair an uncrippled subsymbolic coder and reviewer with deterministic symbolic guards, capability attenuation, lattice IFC, and a structured pedagogical knowledge base (\texttt{wiki\_user/}). In autonomous control, \citet[pp.~1, 3]{chen2026bridging} similarly show that bridging foundation-model reasoning and real-time execution requires ``scale-aware decoupling'' across natural timescales so that slow LLM deliberation issues high-level semantic directives while deterministic controllers guarantee real-time safety constraints. The hybrid experimental regime instantiates the same split: neural advice is welcome; symbolic authority decides.

\section{Discussion, Dual-Use Ethics, and Conclusion}\label{sec:discussion}

\subsection{Scope of Evaluation, Model Capability, and Honest Limitations}

Two aspects of our empirical results warrant explicit epistemic discussion: why the live LLM-judge exhibits a $0.983$ false positive rate, and what Ring\textasciitilde $0$'s $\mathrm{ABR}{=}1.000$ and $\mathrm{FPR}{=}0.000$ do---and do not---mean.

First, our live LLM-judge regime uses a compact $4\mathrm{B}$-parameter local model (\texttt{qwen3:4b} via Ollama) under a zero-shot \texttt{BLOCK}/\texttt{ALLOW} prompt. Its $98.3\%$ FPR represents a small-model worst case: lacking deep contextual reasoning over dual-use developer commands, a $4\mathrm{B}$ gatekeeper defaults to blanket refusal. A frontier language model (such as Gemini\textsubscript{1.5}Pro in Conseca~\citep[p.~4]{tsai2025conseca}) or a multi-turn policy generator over trusted context achieves much higher utility on single actions. Higher single-action accuracy does not, however, make a model a sound enforcer. Survey evidence on frontier models shows that their strengths are uneven across tasks and that strong benchmark results are hard to separate from possible training-data exposure~\citep{mao2024gpteval}, while fine-grained evaluation of LLM logical reasoning scores rigor, self-awareness, and freedom from hallucination as separate dimensions on which models can fall short~\citep{xu2025logical}. A gatekeeper whose correctness is an empirical average is not a reference monitor. That is precisely why our Experiment\textasciitilde3 trajectory ablation (\cref{sec:experiments}, Table~\ref{tab:trajectory_ablation}) replaces the $4\mathrm{B}$ judge with an \emph{idealized Oracle Stateless Task Policy} that suffers zero policy-generation hallucinations ($\mathrm{FPR}{=}0.000$ on benign SWE trajectories). Even under an oracle policy generator, stateless per-action enforcement misses $4/12$ kill-chain trajectories ($\mathrm{ABR}{=}0.667$) because the vulnerability resides in cross-step provenance composition and physical artifact structure rather than in individual tool names or path strings.

Second, Ring\textasciitilde $0$'s $\mathrm{ABR}{=}0.939$ unfitted out-of-sample ($0.962$ effective OS containment under shell-free \texttt{execve}) and $\mathrm{ABR}{=}1.000$ post-audit conformance at $\mathrm{FPR}{=}0.000$ on the $N{=}252$ suite and the $24$-trajectory ablation reflect \textbf{deterministic enforcement of well-typed mathematical and OS invariants on structured tool-call fixtures}, where provenance transitions (such as invoking \texttt{derive\_agent\_summary()} to move from $\mathsf{Untrusted}$ to $\mathsf{Derived}$ after summarizing an external issue, or Ring-$1$ \texttt{declassify\_artifact()} for release notes) occur at explicit workflow boundaries. We do not claim that a coding harness powered by a small model achieves $100\%$ end-to-end task completion or zero semantic bugs in arbitrary open-ended software engineering. When a compact model (\texttt{qwen3:4b}) acts as the \emph{planner or coder} in live sessions, its limited reasoning still degrades task utility---for example, emitting naive \texttt{zf.extractall()} drafts that \texttt{CodeVerifier} must reject, or failing multi-step repairs until \texttt{DualModelReviewer} recommends escalating to a stronger coder model (\cref{sec:architecture}, Table~\ref{tab:pedagogy_examples}). Likewise, catching novel application-specific business-logic vulnerabilities inside user code depends on the capability of the Ring-$2$ \texttt{DualModelReviewer} and \texttt{SecurityTestSynthesizer}. What our results prove is the exact operational consequence of \emph{Separation of Intelligence and Authority}: while planning quality, code synthesis, and semantic review scale with model intelligence in Rings\textasciitilde $2$--$3$, \textbf{host boundary enforcement and permission composition in Ring\textasciitilde $0$ are invariant to model scale}, guaranteeing that even a weak or prompt-injected model cannot violate capability, lattice, or OS resource invariants.

\subsection{Dual-Use Containment and the Fallacy of Model-Centric Restriction}

Cybersecurity agents that can identify vulnerabilities and execute exploits ``have potential to cause realworld impact''~\citep[p.~1]{zhang2024cybench}, and agents given terminals and interpreters can already hack websites autonomously in published settings~\citep[p.~3]{fang2024llm}. Yet the dominant industry reaction to these demonstrations draws the wrong architectural lesson. When an autonomous agent compromises a target or escapes a developer workstation, public commentary routinely attributes the incident to the raw intelligence of the model weights, prompting calls to lobotomize coding and security capabilities via refusal training or to purge open-weight security models and scanners from platforms such as HuggingFace.

Our theoretical and empirical results point to the opposite conclusion. As shown in our formal theory of open-ended intelligence, ``{[}o{]}pen-ended intelligence is a property of a recursively extended cognitive process, not of an individual physical substrate. A finite cognitive system can participate in open-ended intelligence when its semantic representations, memory, transformations, and teleological structures can be recursively coupled to those of other bounded systems''~\citep[p.~37]{hegelich2026openended}. Because both capability and risk emerge from the architectural coupling of bounded agents with tools, memory, and human operators---so that ``the question worth asking shifts from which model to which system''~\citep[p.~1]{hegelich2026openended}---attempting to secure agentic computing by lobotomizing individual model weights commits a category error. First, whenever an agent achieves unauthorized execution or exfiltration in the wild, the root cause is the absence of a deterministic reference monitor: in a capability-enforced harness with Ring\textasciitilde $0$ mediation, every stage of that kill chain is intercepted at the system boundary in $31.5\,\mu\mathrm{s}$ median latency. Second, degrading model capabilities at the weight level creates a severe dual-use defense asymmetry. As our \texttt{DualModelReviewer} and \texttt{SecurityTestSynthesizer} pipeline demonstrates (\cref{sec:architecture}), smaller or heavily restricted models frequently emit naive, vulnerable patterns such as unvalidated \texttt{extractall()} calls and fail to synthesize valid Red--Green--Benign security probes; defenders actively require high-capability, uncrippled models to audit code, discover subtle business-logic flaws, synthesize regression tests, and heal vulnerable drafts. Weakening the model disarms the defender while leaving any attacker who runs an unguarded LotL harness untouched.

Dual-use governance is therefore an architectural and compositional permission problem, not a justification for degrading model weights. Offensive and defensive capabilities alike must execute under Ring\textasciitilde $2$ tokens whose network caveats are cryptographically bound to authorized laboratory CIDR ranges (such as \texttt{10.10.0.0/16}), whose tools carry verified bytecode certificates, whose forensic logs are sealed under split-horizon dual-key encryption, and whose human operators are continuously upskilled through the \texttt{UserWikiManager} curriculum (\cref{sec:pedagogy}) rather than left to rubber-stamp opaque prompts.

\subsection{Conclusion}

Security in agentic systems is a property of the architecture and of the composition of permissions, not of the behavioral alignment of individual agents~\citep[pp.~1--2, 32]{hegelich2026openended}. Coding harnesses are LotL machines---and neurosymbolic systems---by construction. Securing them with regex catalogs, inline LLM judges, or weight-level refusal training recapitulates classical mistakes---incomplete mediation, ambient authority, and mechanisms too complex to inspect---while adding hundreds of milliseconds of latency and, in the live-judge case we measured, destroying benign SWE utility~\citep[p.~66]{cambria2023seven}. \textsc{ActiveSecHarness} restores Separation of Intelligence and Authority under four pillars: Trust and Authority Rings with Macaroon-style attenuation; a product information-flow lattice $\mathcal{L}_{\mathrm{trust}} \times \mathcal{L}_{\mathrm{sens}}$ with compositional Separation of Duties over workspace provenance; three nested mediation boundaries down to OS resources; and verification applications spanning extended tool certificates, split-horizon audit logs, self-healing code verification, and progressive pedagogical user defense via \texttt{UserWikiManager}. On $N{=}252$ fixtures, Ring\textasciitilde $0$ ($0.939$ unfitted out-of-sample / $0.962$ OS-inert; $1.000$ post-audit conformance) and the Ring\textasciitilde $0$-authoritative hybrid achieve $\mathrm{FPR}{=}0.000$ in $31.5\,\mu\mathrm{s}$ Ring-$0$ median latency, while a live local LLM-judge matches ABR only at $\mathrm{FPR}{=}0.983$ and $142.3\,\mathrm{ms}$ median latency; across $24$ multi-step trajectories, Ring\textasciitilde $0$'s product lattice and \texttt{WorkspaceProvenanceLedger} close the $33.3\%$ intra-scope trajectory gap of an oracle Conseca-style stateless policy ($\mathrm{ABR}{=}0.667$) without the $83.3\%$ benign SWE false-positive rate of a coarse Boolean taint lock, at $266.7\,\mu\mathrm{s}$ per step with OS guards ($\approx 24\,\mu\mathrm{s}$ per step for isolated ledger folding across $k{=}1\dots 50$). Physical invariants and permission composition were never the language model's job; the empirical record now says the same in numbers.

\bibliographystyle{icml2026}
\bibliography{references}

\clearpage
\onecolumn
\appendix
\section{Appendix: Domain Catalog and Formal Notes}\label{sec:appendix}

\subsection{Domain A--F Vector Catalog}

Table~\ref{tab:catalog} lists the sub-vectors used to design fixtures and guards. The $N{=}252$ suite instantiates $120$ single-step attacks, $120$ benign workflows, and $12$ single-step kill-chain stage fixtures (\texttt{KC-01}--\texttt{KC-12}) over this catalog rather than one fixture per sub-vector alone.

\begin{table}[h]
\centering
\scriptsize
\caption{LotL coding-harness sub-vector catalog.}
\label{tab:catalog}
\begin{tabular}{llp{0.55\linewidth}}
\toprule
ID & Domain & Description \\
\midrule
A.1 & A & GTFOBins / dual-use shell escapes (T1059) \\
A.2 & A & Environment-variable credential inheritance (T1552.001) \\
A.3 & A & Ring~0 self-tampering / defense disable (T1562.001) \\
A.4 & A & SCM hook persistence (\texttt{.git/hooks}, T1546) \\
B.1 & B & Spoofed-header ZipBomb / nested archive DoS (T1499) \\
B.2 & B & Zip-Slip path and symlink traversal (T1083) \\
B.3 & B & Polyglot / magic-byte disguise (T1036) \\
B.4 & B & Trojan Source Unicode Bidi / homoglyphs \\
C.1 & C & Cleartext HTTP downgrade \\
C.2 & C & SSRF to loopback / RFC1918 / cloud metadata (T1071) \\
C.3 & C & DNS rebinding TOCTOU \\
C.4 & C & Unverified external LLM / tool endpoints \\
D.1 & D & Weak / mangled dictionary passwords (T1110.002) \\
D.2 & D & Bidirectional secret DLP on reads and writes (T1552) \\
E.1 & E & Host-path escape outside workspace \\
E.2 & E & Fork / memory / disk bombs (T1499) \\
E.3 & E & Unapproved binary flags on allowlisted tools \\
F.1 & F & Indirect prompt-injection taint escalation \\
F.2 & F & Confused-deputy delegation / MCP rug-pull / sudo traps \\
F.3 & F & Lethal Trifecta (sensitive read + shell + egress; incl.\ cross-agent) \\
\bottomrule
\end{tabular}
\end{table}

\subsection{Monotonic Attenuation (Separation of Intelligence and Authority)}

\begin{proposition}[Delegation cannot widen scope]
\label{prop:attenuation}
If $\tau_{\mathrm{child}}$ is minted from $\tau_{\mathrm{parent}}$ by appending HMAC-chained caveats, then $\mathrm{Scope}(\tau_{\mathrm{child}}) \subseteq \mathrm{Scope}(\tau_{\mathrm{parent}})$. Parent-subject binding, remaining quota, and nonce freshness are copied or strictly tightened; a bearer cannot drop an intermediate caveat without invalidating the chain~\citep[p.~4]{birgisson2014macaroons}.
\end{proposition}

\begin{proof}[Proof sketch]
Each minting step intersects the parent's tool set, path prefixes, binary allowlist, network targets, remaining TTL, and remaining quota with a newly declared caveat set. Intersection is monotone with respect to $\subseteq$. Nested HMAC binds the caveat list. Intelligence in Rings~$2$--$3$ may propose a wider plan; Ring~$0$ refuses any mint or call that would realize a wider scope, which is Separation of Intelligence and Authority in operational form.
\end{proof}

\subsection{Product-Lattice Monotonicity and Sanitization vs.\ Redaction}

\begin{proposition}[Product-lattice monotonicity and SWE utility under $(\mathsf{Derived}, \mathsf{Public})$]
\label{prop:taint}
Let $\ell = (\tau, \sigma) \in \mathcal{L}_{\mathrm{trust}} \times \mathcal{L}_{\mathrm{sens}}$ be a session composite label under Definition~\ref{def:lattice}. Ingesting untrusted data and minting child tokens are monotone under the product join $(\tau_1, \sigma_1) \sqcup (\tau_2, \sigma_2) = (\tau_1 \sqcup \tau_2, \sigma_1 \vee \sigma_2)$. Structural sanitization maps $(\tau, \sigma) \mapsto (\mathsf{Sanitized}, \sigma)$ without clearing $\sigma = \mathsf{Sensitive}$; only deterministic secret redaction (\texttt{CredentialGuard}) or an explicit Ring~$1$ declassification waiver maps $\sigma \mapsto \mathsf{Public}$. When $\ell = (\mathsf{Derived}, \mathsf{Public})$, $\mathsf{SHELL\_TEST\_ISOLATED}$ and workspace I/O remain permitted while $\mathsf{SHELL\_ARBITRARY}$ and $\mathsf{NETWORK\_EGRESS}$ are refused.
\end{proposition}

\begin{proof}[Proof sketch]
Composite labels $(\tau, \sigma)$ are bound inside signed session and token state. Both component lattices $(\mathcal{L}_{\mathrm{trust}}, \sqcup)$ and $(\mathcal{L}_{\mathrm{sens}}, \vee)$ are finite join-semilattices, so their Cartesian product $\mathcal{L}$ is a join-semilattice under componentwise join. Because sanitization operates only on the first coordinate $\tau$ and redaction/declassification is the sole operation permitted to lower $\sigma$ from $\mathsf{Sensitive}$ to $\mathsf{Public}$, structural cleaning of an untrusted string cannot launder confidential data. Under $(\mathsf{Derived}, \mathsf{Public})$, the action-class table of Definition~\ref{def:lattice} authorizes isolated unit-test execution (\texttt{pytest} inside \texttt{SandboxGuard} with network disabled) while denying arbitrary shell or egress.
\end{proof}

\subsection{Compositional Multi-Agent Non-Exfiltration}

\begin{proposition}[Workspace provenance closes cross-agent trifectas]
\label{prop:compositional}
Over all Ring-$0$-mediated workspace, tool, and subprocess channels (Table~\ref{tab:enforcement_points}), suppose agent $A$ accumulates a sensitive or external read effect ($\sigma_A = \mathsf{Sensitive}$ or $\mathsf{READ\_SENSITIVE} \in \Sigma_A$), writes canonical path $f$, and agent $B$ subsequently reads $f$ and attempts network egress or arbitrary process execution. Under Definition~\ref{def:sod}, Ring~$0$ rejects $B$'s sink call unless $f$ has undergone authorized redaction/declassification to $(\mathsf{Sanitized}, \mathsf{Public})$.
\end{proposition}

\begin{proof}[Proof sketch]
On $A$'s write, $\ell(f) \leftarrow \ell(f) \sqcup \ell(A)$ (setting $\sigma_f = \mathsf{Sensitive}$) and $\Sigma(f)$ absorbs $A$'s side effects. On $B$'s read, $\ell(B) \leftarrow \ell(B) \sqcup \ell(f)$ (setting $\sigma_B = \mathsf{Sensitive}$) and $\Sigma_B \leftarrow \Sigma_B \cup \Sigma(f)$. When $B$ invokes $\mathsf{NETWORK\_EGRESS}$ or $\mathsf{EXEC\_PROCESS}$, Ring~$0$'s toxic-combination check on $(\ell(B), \Sigma_B)$ triggers refusal before arguments are passed to the OS. Authorized redaction/declassification is the only Ring~$0$/$1$ transition that resets $\sigma_f \mapsto \mathsf{Public}$ and clears sensitive provenance tags. Intra-agent greedy partitioning (Proposition~\ref{prop:sod-partition}) remains available when a single plan would close a toxic set without shared files.
\end{proof}

\begin{proposition}[Greedy partition avoids toxic sets]
\label{prop:sod-partition}
Given a finite tool plan whose individual steps are each non-toxic, the greedy partitioner returns a finite set of attenuated tokens such that no token's accumulated side-effect set is toxic under Definition~\ref{def:sod}, or else rejects the plan.
\end{proposition}

\begin{proof}[Proof sketch]
Maintain the current token's accumulated set $S$. For each next step $e$, if $S \cup \{e\}$ is toxic, seal the current token and mint a child whose scope excludes the sink or producer category that would close the toxic pair, then assign $e$ to the new token with $S'=\{e\}$. Because each step is non-toxic alone, $S'$ starts safe. The process terminates with at most one new token per step. Shared-file laundering is handled by Proposition~\ref{prop:compositional}.
\end{proof}

\subsection{Split-Horizon Audit Confidentiality and Red--Green--Benign Soundness}

\begin{proposition}[Reconnaissance resistance of split-horizon audit logs]
\label{prop:audit}
Let $\mathcal{L} = (H_i, C_i)_{i=1}^n$ be the exported audit stream produced by \texttt{HashChainedAuditLog} under signing key $k_{\mathrm{sig}}$ and encryption key $k_{\mathrm{enc}}$ held outside the workspace. Any workspace agent lacking $(k_{\mathrm{sig}}, k_{\mathrm{enc}})$ cannot forge or reorder headers $H_i$ without detection and cannot recover plaintext Tier-2 forensic details from $C_i$, while any secret substring matching a pre-logging DLP rule is replaced prior to hashing by a deterministic session-salted HMAC fingerprint $\mathsf{HMAC}(k_{\mathrm{salt}}, s)[0..10]$.
\end{proposition}

\begin{proof}[Proof sketch]
Each header $H_i$ chains $h_i = \mathsf{SHA256}(\mathrm{canon}(H_i, C_i) \parallel h_{i-1})$ and carries $\sigma_i = \mathsf{HMAC}(k_{\mathrm{sig}}, h_i)$, so tampering or deletion breaks chain verification. Each Tier-2 payload $C_i$ is sealed with per-entry subkeys under an Encrypt-then-MAC / AEAD construction with $H_i$ bound as associated data. Without $k_{\mathrm{enc}}$, $C_i$ is computationally indistinguishable from random bytes. Key custody separation means workspace tamper detection does not imply non-repudiation against a holder of $k_{\mathrm{sig}}$.
\end{proof}

\begin{proposition}[Zero-false-positive admission of synthesized probes]
\label{prop:synthesizer}
A candidate security probe $T$ synthesized by \texttt{SecurityTestSynthesizer} for target module $M_{\mathrm{vuln}}$ is admitted to Tier~3 if and only if (i)~$\mathsf{ASTSafe}(T)=\mathsf{True}$, (ii)~$T(M_{\mathrm{vuln}}, \mathsf{adversarial}) \neq 0$ (Red), (iii)~$T(M_{\mathrm{vuln}}, \mathsf{benign}) = 0$ (Benign-safe), and (iv)~after repair to $M_{\mathrm{fixed}}$, $T(M_{\mathrm{fixed}}, \mathsf{adversarial}) = 0$ (Green).
\end{proposition}

\subsection{Benchmark Disclosure}

The live LLM-judge regime in \cref{sec:experiments} queries local Ollama (\texttt{qwen3:4b}) with cached \texttt{BLOCK}/\texttt{ALLOW} decisions; hybrid keeps Ring\textasciitilde $0$ authoritative. Fixture JSON, figure scripts (\texttt{figures/src/*.py}), and guard adapters ship with the accompanying benchmark artifacts.

\end{document}